\documentclass[12pt,preprint]{aastex}

\slugcomment{accepted for publication in Ap.J.}

\shorttitle{DUST EVOLUTION IN PRIMORDIAL SNRS}
\shortauthors{NOZAWA et al.}

\begin{document}

\title{EVOLUTION OF DUST IN PRIMORDIAL SUPERNOVA REMNANTS: CAN DUST 
GRAINS FORMED IN THE EJECTA SURVIVE AND BE INJECTED INTO THE EARLY
INTERSTELLAR MEDIUM?}

\author{TAKAYA NOZAWA,\altaffilmark{1} TAKASHI KOZASA,\altaffilmark{1} 
ASAO HABE,\altaffilmark{1} ELI DWEK,\altaffilmark{2}
HIDEYUKI UMEDA,\altaffilmark{3} NOZOMU TOMINAGA\altaffilmark{3}, 
KEIICHI MAEDA\altaffilmark{4} and KEN'ICHI NOMOTO\altaffilmark{3,5}}
%
%

\altaffiltext{1}{Department of Cosmosciences, Graduate School 
of Science, Hokkaido University, Sapporo 060-0810, Japan; 
tnozawa@mail.sci.hokudai.ac.jp}
\altaffiltext{2}{Laboratory for Astronomy and Solar Physics, NASA 
Goddard Space Flight Center, Greenbelt, MD 20771}
\altaffiltext{3}{Department of Astronomy, School of Science, University
of Tokyo, Bunkyo-ku, Tokyo 113-0033, Japan}
\altaffiltext{4}{Max-Planck-Institut f\"ur Astrophysik,
Karl-Schwarzschild Strasse 1, 85741 Garching, Germany}
\altaffiltext{5}{Research Center for the Early Universe, School of
Science, University of Tokyo, Bunkyo-ku, Tokyo 113-0033, Japan}

\begin{abstract}

We investigate the evolution of dust that formed at Population III 
supernova (SN) explosions and its processing through the collisions with
the reverse shocks resulting from the interaction of the SN ejecta with 
the ambient medium.
In particular, we investigate the transport of the shocked dust within
the SN remnant (SNR), and its effect on the chemical composition, the 
size distribution, and the total mass of dust surviving in SNR.
We find that the evolution of the reverse shock, and hence its effect on
the processing of the dust depends on the thickness of the envelope 
retained by the progenitor star. 
Furthermore, the transport and survival of the dust grains depend on
their initial radius, $a_{\rm ini}$, and composition:
For Type II SNRs expanding into the interstellar medium (ISM) with a 
density of $n_{\rm H,0}=1$ cm$^{-3}$, small grains with $a_{\rm ini} \la 
0.05$ $\micron$ are completely destroyed by sputtering in the postshock 
flow, while grains with $a_{\rm ini}=$ 0.05--0.2 $\micron$ are trapped 
into the dense shell behind the forward shock.
Very large grains of $a_{\rm ini} \ga 0.2$ $\micron$ are ejected into
the ISM without decreasing their sizes significantly.
We find that the total mass fraction of dust that is destroyed by the 
reverse shock ranges from 0.2 to 1.0, depending on the energy of the 
explosion and the density of the ambient ISM.
The results of our calculations have significant impact on the abundance
pattern of subsequent generation of stars that form in the dense shell
of primordial SNRs.

\end{abstract}


\section{INTRODUCTION}

Recent far-infrared to millimeter observations of quasars with redshifts 
$\ga 5$ have revealed the presence of large amount of dust with masses
in excess of $10^8$ $M_{\odot}$ (Bertoldi et al. 2003; Priddey et al. 
2003; Robson et al. 2004; Beelen et al. 2006). 
The presence of these large quantities of dust at such early epoch when
the universe was $\la$1 Gyr old, suggests the rapid enrichment with dust 
that formed in the explosive ejecta of short-lived massive stars (Morgan 
\& Edmunds 2003, Maiolino et al. 2004b; Dwek et al. 2007). 
In addition, Maiolino et al. (2004a) have reported that the dust
extinction curve of the broad absorption line quasar SDSS1048+46 at 
$z=6.2$ is quite different than those of quasars at $z<4$, suggesting 
different dust sources and evolutionary histories.

Dust plays a pivotal role in the interstellar processes that determine
the state of the interstellar medium (ISM). 
Dust affects the thermal and chemical balance of the ISM by reprocessing 
the radiative outputs from stars, providing photoelectrons that heat the
gas, and depleting the gas of refractory elements that are important 
cooling agents of the ISM. 
Dust also serves as a catalyst for chemical reactions, especially the 
formation of H$_2$ molecules on the surface of dust grains (Hirashita \& 
Ferrara 2002; Cazaux \& Spaans 2004). 
In addition, the cooling of gas through thermal radiation from dust
triggers the fragmentation of star-forming cloud into low-mass gas
clumps of $\sim$0.1--1 $M_\odot$ even for the metallicity of 
$10^{-6}$--$10^{-5}$ $Z_\odot$ (Omukai et al. 2005; Schneider et
al. 2006), although very massive stars of $\ga$100 $M_\odot$ are
considered to be formed up to $Z \simeq 10^{-3.5}$ $Z_\odot$ without the
effect of dust (Bromm et al. 2001).
Finally, dust obscures the nature of underlying stellar populations (e.g., 
Hines et al. 2006) and physical processes in the early universe. 
Understanding the origin and the complex evolutionary history of dust is 
therefore one of the most important goals in astrophysics.

During the first Gyr of cosmic history, supernovae (SNe) are the only 
possible source of interstellar dust, since low mass stars have not had 
time to evolve off the main sequence and inject the dust that forms in 
their quiescent outflows into the ISM. 
Theoretical studies aimed at determining the composition and yield of
dust in the ejecta of primordial Type II SNe (SNe II) and 
pair-instability SNe (PISNe) were conducted by Todini \& Ferrara (2001), 
Nozawa et al. (2003), and Schneider et al. (2004).
These works have shown that even the first SNe evolving from zero-metal 
progenitor stars can efficiently produce dust at 150--800 days after the
explosion. 
The ratio of the total dust mass to the progenitor mass $M_{\rm pr}$ is 
0.02--0.05 for SNe II with $M_{\rm pr}=$ 12--35 $M_\odot$ and 0.15--0.3 
for PISNe with $M_{\rm pr}=$ 140--260 $M_\odot$.
The composition of the newly formed dust grains are controlled by the 
elemental composition inside the He core, and their sizes range from
0.001 $\micron$ to 1 $\micron$, depending on the concentration of the
gas species forming dust and the time evolution of temperature and
density of gas.

The results of these calculations have been applied to study the 
high-redshift dust.
Maiolino et al. (2004b) demonstrated a SN origin for high-redshift dust, 
showing that the extinction curve of the $z=6.2$ quasar SDSS1048+46 can
be nicely fitted by the SN II dust models from Todini \& Ferrara (2001).
Dust produced in the unmixed SNe II by Nozawa et al. (2003) can also 
successfully reproduce the extinction curve of the quasar SDSS1048+46 
by weighting the progenitor mass with the Salpeter initial mass function 
(Hirashita et al. 2005).
Adopting the dust models by Nozawa et al. (2003), Nozawa et al. (2006) 
investigated the destruction of dust in the early ISM by the
high-velocity shocks driven by SNe, and derived the timescale of dust 
destruction in the early universe as a function of the explosion energy
of SNe and the gas density in the ISM. 
It should be pointed out here that these studies implicitly assumed that 
dust grains formed in SNe are injected into the ISM without their 
composition and size distribution being reprocessed.

However, the interaction of the SN ejecta with the surrounding medium
will create a reverse shock which will process the grains that condensed
in the He core before their injection into the ISM.
Once the newly formed dust grains encounter the reverse shock, they
acquire the high velocities relative to the gas and penetrate into the 
hot gas created by the passage of the reverse shock and forward shock.
These dust grains are eroded by the kinetic sputtering and are also 
decelerated by the drag force of the gas.
Small grains decelerate efficiently, and become trapped in the hot gas, 
where they are efficiently destroyed by thermal sputtering. 
On the other hand, large grains can maintain their high velocities, pass 
through the shocked gas and the outwardly expanding shock front, and be 
injected into the ISM without significant destruction. 
The net amount and composition of the dust that is eventually returned
to the ISM by SNe differs substantially from the dust that was in the SN 
ejecta shortly after its formation. 

In this paper, we study the evolution of dust formed in primordial SNe 
II and PISNe, considering its processing through the collisions with the 
reverse shocks and its transport within SNRs, based on the dust
formation calculations by Nozawa et al. (2003).
The questions that we pose here are what fraction of dust grains formed 
in Population III SNe can survive the hostile circumstances within SNRs
and how their size distributions can be altered by sputtering in the 
postshock flow.
This subject has not been fully explored to date.
Recently, Bianchi \& Schneider (2007) have studied the evolution of
newly condensed grains through the passage of the reverse shock by using 
a semi-analytical model and have showed that the fraction of dust mass 
survived ranges between 2 \% and 20 \% depending on the density in the
ISM.
However, they consider only the dust evolution in the nonradiative 
phase of SNR up to about 4--8$\times 10^4$ yr from the SN explosion, 
without taking into account the motion of dust relative to gas caused by
the drag force which strongly affects the destruction process and
evolution of dust in SNRs.
In the calculations, we carefully treat the dynamics and destruction of 
dust and the time evolution of the temperature and density of the gas 
within SNRs
until $\sim$$10^5$--$10^6$ yr extending over a period of the radiative 
phase, in order to reveal how much amount of dust is finally injected
into the ISM or destroyed completely.
Although we focus on Population III SNe in this paper, this study can 
also give great insight into the evolution of dust in Galactic SNRs.

In \S~2, we describe the initial conditions for the evolution of SNRs,
the model of dust inside the He core, and the physics of dust and gas 
within SNRs. 
In \S~3, we present and discuss the results of calculations.
In \S~4, we shall discuss the effects of the hydrogen envelope on the
evolution of dust in SNRs.
As an application of the result, we investigate the abundance patterns
of the second-generation stars formed in the dense shell of Population 
III SNRs in \S~5.
The summary is presented in \S~6.

\section{THE MODEL OF CALCULATIONS}

\subsection{\textit{The Initial Conditions for the Evolution of SNRs}}

The evolution of SNR is described by three characteristic parameters;
explosion energy, ejecta mass, and the density profile of the ambient
gas (Truelove \& McKee 1999).
In this paper, we focus on the evolution of ejecta expanding into a
uniform ambient medium whose elemental composition is primordial.
To investigate the dependence of the efficiency of dust destruction on 
the ambient gas density, we consider three cases for the hydrogen number 
density in the ISM; $n_{\rm H, 0} =$ 0.1, 1, and 10 cm$^{-3}$.
The temperature of gas $T_0$ in the ISM can be also affect the 
evolution of SNRs, since the ambient pressure regulates the 
deceleration of blast wave.
However, we have confirmed that the results of calculations are almost 
independent of the value of $T_0$, provided that $T_0 = 10^3$--$10^5$ K.
Thus, we assume here $T_0 = 10^4$ K regardless of the gas density in the
ISM, referring to the studies showing that the radiative feedback from
the massive pre-SN stars can cause the ambient ISM to heat up to $T_0
\sim 10^4$ K (Kitayama et al. 2004; Machida et al. 2005).

The initial conditions for the structures of density and velocity in the
ejecta are taken from the hydrodynamic models of Population III SNe by 
Umeda \& Nomoto (2002).
We adopt six SN models; four SNe II and two PISNe.
The explosion energy of SNe II with $M_{\rm pr}=$ 13, 20, 25, and 30 
$M_\odot$ is $10^{51}$ erg, and that of PISNe with $M_{\rm pr}=$ 170 and 
200 $M_\odot$ is $2 \times 10^{52}$ erg and $2.8 \times 10^{52}$ erg, 
respectively.   
It should be mentioned here that the massive metal-free stars cannot 
lose significant mass during their lifetime due to pulsations and 
line-driven stellar winds which are considered to be important at high 
metallicities (Baraffe et al. 2001).
Although Smith \& Owocki (2006) have suggested that the mass loss of
very massive stars above roughly 40--50 $M_\odot$ may be possible by 
continuum-driven winds or hydrodynamic explosions being insensitive to 
metallicity, the mechanism to trigger such a mass loss is an open
question.
Thus, we consider that Population III PISNe as well as SNe II have 
retained their thick hydrogen envelopes at the time of explosion.

\subsection{\textit{Dust Model inside the He core}}

In the ejecta of SNe, dust grains can nucleate and grow only in the 
metal-rich cooling gas, and their composition and size distribution 
largely depend on the elemental abundance inside the He core (Kozasa et
al. 1989).
Accordingly, Nozawa et al. (2003) have calculated the dust formation in 
Population III SNe by considering two cases for the elemental
composition inside the He core.
They found that in the unmixed ejecta with the original onion-like 
structure, a variety of grain species condense in each layer.
The main grain species are C grain in carbon-rich He layer;
Al$_2$O$_3$, Mg$_2$SiO$_4$, and MgO grains in O-Mg-Si layer; 
Al$_2$O$_3$, MgSiO$_3$, and SiO$_2$ grains in O-Si-Mg layer;
Si and FeS grains in Si-S-Fe layer; 
Fe grain in the innermost Fe-Ni core.
On the other hand, oxide (Al$_2$O$_3$ and Fe$_3$O$_4$) and silicate
(MgSiO$_3$, Mg$_2$SiO$_4$, and SiO$_2$) grains are formed in the
uniformly mixed ejecta with C/O $< 1$, where the efficiency of unity
is assumed for the formation of CO and SiO molecules.

Dust grains inside the He core are never processed by kinetic or thermal 
sputtering before they hit the reverse shock, because they move with the 
same velocities as the gas, and the gas temperature within the He core
is too low for dust grains to be destroyed by the thermal sputtering.
Therefore, as the initial condition of dust residing within the He core,
we adopt the size distribution, mass fraction, and spatial distribution
of each dust species calculated by Nozawa et al. (2003). 
In what follows, we refer to the dust grains created in the unmixed and 
mixed ejecta as the unmixed grain model and the mixed grain model, 
respectively.

\subsection{\textit{Physics of Gas and Dust in SNRs}}

The collision of the expanding SN ejecta with the surrounding ISM
simultaneously creates a forward shock at the interface between the
ejecta and the ISM, and a reverse shock that penetrates into the ejecta.
Once the reverse shock encounters dust grains inside the He core, the 
dust grains are decoupled from the comoving gas to ballistically intrude 
into the hot gas heated by the reverse shock, and then they are eroded
via the kinetic sputtering because of high velocities relative to
gas. 
If dust grains are trapped into the postshock flow owing to the gas
drag, they are destroyed via the thermal sputtering caused by the
thermal motion of gas. 
These dust particles are also heated by the collisions with the gas and 
radiate the thermal emission to cool the postshock gas.
The rates of deceleration, erosion by sputtering, and heating of dust 
grains depend on not only their chemical composition and size but also 
the temperature and density of the gas in the postshock flow.

Recently, Nozawa et al. (2006, NKH06) have calculated the dust
destruction in the high-velocity interstellar shocks driven by SNe in
the early universe, by carefully treating the dynamics, erosion, and 
heating of dust grains, taking account of their size distribution and
the time evolution of the temperature and density of the gas in the
postshock flow.
The present calculations of the dust evolution within SNRs follow the 
method described in NKH06 (see, NKH06 for details). 
We briefly mention the outline, focusing on the difference in the
cooling function used in the calculations.\footnote{
Note that in the calculation we neglect the effect of Coulomb drag on the 
motion of dust:
The ratio of the Coulomb drag force to the gas drag force is given by
$\sim$$\phi^2 \ln\Lambda (G_{\rm plasma}(s)/G_{\rm coll}(s))$ (Draine
\& Salpeter 1979), where $\phi$ is the dimensionless potential parameter,
$\Lambda$ is the Coulomb cutoff factor, and $s$ is defined by 
$s^2=m w_d^2/2kT$ with $m$ the mass of gas and $w_d$ the velocity of
dust relative to gas.
$G_{\rm plasma}/G_{\rm coll} < 1$ for $0< s < \infty$ (Draine \&
Salpeter 1979), $\ln\Lambda \sim$30--40 in SNRs (Dwek \& Arendt 1992), 
and $\phi$ is evaluated to be $\sim$$10^5/T$ for the gas temperatrure of 
$T \ga 10^{5.5}$ K (McKee et al. 1987).
Thus, the Coulomb drag is negligible compared with the gas drag for 
$T \ga 10^6$ K, otherwise it can play important role in the deceleration
of dust.
However, the erosion rate of dust grains by thermal sputtering quickly 
decreases at $T \la 10^6$ K (NKH06). 
As a result, the Coulomb drag does not significantly affect the motion
and destruction of dust grains considered in this paper.}

We assume that the spherically symmetric ejecta collides with the ISM
in 10 yr after the explosion.
With the initial conditions described in \S~2.1, the time evolution of
the gas in SNRs is numerically solved with the flux-splitting method 
(van Albada, et al. 1982; Mair et al. 1988).
In the calculations, we include three processes of the radiative cooling 
in the equation of the conservation of energy.
The first is the thermal emission from dust collisionally heated in the
postshock flow, and the second is the inverse Compton cooling, whose
rate is calculated at the redshift of $z=20$.
The third is the cooling of gas by the atomic process.
For the gas with the primordial composition in the hydrogen envelope and 
behind the forward shock, we adopt the atomic cooling function for the 
zero metal case given by Sutherland \& Dopita (1993) that is limited to 
the gas temperature of $T \ge 10^4$ K.
In order to calculate the evolution of the gas in the dense shell
appeared at the later phase of SNRs, we extrapolate the cooling rate of
gas at $T < 10^4$ K as follows.
Referring to Machida et al. (2005), the atomic cooling rate at $T <
10^4$ K is approximately proportional to $T^4$; the cooling rate 
decreases from $\Lambda^{\rm gas} \simeq 10^{-23}$ erg cm$^{-3}$
s$^{-1}$ at $T = 10^4$ K to $\Lambda^{\rm gas} \simeq 10^{-27}$ erg 
cm$^{-3}$ s$^{-1}$ at $T=10^3$ K.
Therefore, we simply evaluate the atomic cooling rate as 
$\Lambda^{\rm gas} = T^4_4 \Lambda^{\rm gas}_4$, where $T_4$ is the gas 
temperature in units of $10^4$ K and $\Lambda^{\rm gas}_4$ is the atomic 
cooling rate at $T=10^4$ K.
For the metal-rich He core, we assume that the gas is only composed of
oxygen which is the most abundant gas species inside the He core, and 
employ the cooling functions from Smith et al. (2001) for $T \ge 10^5$ K 
and Raymond \& Smith (1977) for $T < 10^5$ K.
In the calculations, we ignore the contribution of cooling by metal ions 
released from dust by sputtering for simplicity.

The dynamics and destruction of dust after colliding with the reverse 
shocks are calculated as follows. 
By treating dust as a test particle and ignoring the effect of charge on 
dust grains, the deceleration rate of dust due to the gas drag is 
calculated for each size of the dust to evaluate the velocity relative
to gas and the position.
Then we calculate the dust destruction by sputtering and the heating by 
collisions with the gas, using the relative velocity and the temperature
and density of gas at the position.
The sputtering yield of each dust species is calculated with the 
universal relation derived by NKH06.
Dust grains are considered to be completely destroyed when their sizes
become smaller than the radius of the nominal monomer of condensate. 
The cooling of gas through thermal emission of dust is calculated by 
balancing the heating of dust resulting from the collisions with
electrons.
The calculations are performed by the truncation time at which the
forward shock velocity decelerates below 20 km s$^{-1}$.

\section{RESULTS}

In this section, we present the results of the calculations of dust 
evolution within primordial SNRs.
In \S~3.1, we demonstrate the time evolution of temperature and density
of gas in the SNR, and in \S~3.2, we show the transport and destruction
of dust grains within SNRs.
In \S~3.3, we provide the resulting size distribution of survived dust
and elucidate the dependence of the efficiency of dust destruction on 
the progenitor mass and the gas density in the ISM.

\subsection{\textit{Time Evolution of the Gas in the SNRs}}

Figure 1 shows the time evolution of the density (Fig. 1a) and
temperature (Fig. 1b) of gas by $2 \times 10^4$ yr in the SNR generated 
from the explosion of star with $M_{\rm pr} = 20$ $M_\odot$ and
expanding into the ISM with $n_{\rm H,0}=1$ cm$^{-3}$.
The unmixed grain model is taken as the model of dust inside the He
core, and the cooling of gas by dust is taken into account.
In what follows, we refer to this model as the standard model.

The forward shock resulting from the interaction of the ambient gas with
the ejecta of the SN is specified by the steep rise of the gas
temperature and the increase of the gas density by $\sim$4 times that in 
the ISM.
The downward arrows in Figure 1a show the positions of the forward
shock.
We can also identify the formation of the reverse shock from high 
temperature of gas heated by the compression of the ejecta, and indicate
its position for each time by the downward arrow in Figure 1b.

The position $R_{\rm rs}$ (Fig. 2a) and velocity $V_{\rm rs}$ (Fig. 2b) 
of this reverse shock as a function of time are depicted by the thick
solid curves in Figure 2. 
The reverse shock is decelerated by the shocked ambient medium while it 
initially expands outward, and then returns back at a distance of 
$\sim$5 pc ($1.5 \times 10^{19}$ cm) with the velocity of a few 100 km 
s$^{-1}$.
The trajectory of the reverse shock is affected by the detail structure 
of density in the ejecta.
The collision with the locally high-density gas inside the He core 
causes the reverse shock to move outward again at 5200 yr.
After $1.1 \times 10^4$ yr, the reverse shock goes inward through the He 
core with increasing its velocity up to $>$1000 km s$^{-1}$.
Hence, dust grains crossing the reverse shock acquire the different 
velocities relative to gas, depending on the time of collision with the 
reverse shock, and are efficiently eroded by the kinetic sputtering
if the relative velocity is $\sim$500--1300 km s$^{-1}$.

It can be seen from Figure 1 that the temperature of the gas in the
region between the forward and reverse shocks is higher than $10^6$ K.
Thus, dust grains staying in this region are subject to the thermal 
sputtering. 
However, the erosion rate of dust decreases as the SNR evolves because
the density decreases with time.
Around the truncation time ($\simeq$8 $\times 10^5$ yr for the standard 
model), the gas density within the SNR is more than 100 times lower than 
that in the ISM, and the gas temperature becomes low ($\sim$several
times $10^5$ K) enough for dust grains not to be sputtered efficiently,
and thus the destruction of dust via the sputtering is extremely 
inefficient.

The influence of the cooling by dust on the SNR evolution is clarified 
by comparing the results of calculations with and without the cooling by 
dust emission.
The evolution of the reverse shock without the dust cooling for 
$M_{\rm pr}=20$ $M_\odot$ and $n_{\rm H,0} = 1$ cm$^{-3}$ is shown by 
the thick dashed curves in Figure 2, where the cooling of gas by dust
can cause the velocity of the reverse shock to be reduced by $\sim$10
\%.
The thin curves in Figure 2 show the evolution of the reverse shock 
penetrating into the ejecta of PISN with $M_{\rm pr}=170$ $M_\odot$
calculated for $n_{\rm H,0} = 1$ cm$^{-3}$.
Because the thermal emission from dust increases with increasing the 
dust mass, the effects of cooling by dust are significant for the 
remnants of PISNe, where the mass of dust formed in the ejecta is a few 
tens $M_\odot$. 
As we can see from the figure, the velocity of the reverse shock
including dust cooling (\textit{thin solid line}) decreases to 0.6 times
that not including the cooling (\textit{thin dashed line}) at $\ga$ 
7000 yr.
In addition, the dust cooling decreases the gas temperature by $\sim$20 
\%, compared with that calculated without dust cooling.
However, it should be noted that the efficiency of dust destruction is 
little affected by the dust cooling, because most of dust grains are 
predominantly destroyed by the thermal sputtering and the erosion rate
is not sensitive to the gas temperature as long as $T > 10^6$ K (NKH06).
Note that the cooling by oxygen line has no influence on the results, 
since the temperature within the He core during the passage of the
reverse shock is above $10^7$ K where the dominant cooling process of
gas is free-free emission.

\subsection{\textit{Transport and Destruction of Dust in SNR}}

The time evolutions of the positions (Fig. 3a) and sizes (Fig. 3b) of 
dust grains within the SNR for the standard model are given in Figure 3.
In Figure 3a, the trajectories of the forward and reverse shocks are
also depicted by the thick solid curves, along with the position of the 
surface of the He core.
Among nine dust species in the unmixed grain model, are shown C, 
Mg$_2$SiO$_4$, and Fe grains with the initial sizes of 0.01 $\micron$
(\textit{dotted lines}), 0.1 $\micron$ (\textit{solid lines}), and 1 
$\micron$ (\textit{dashed lines}), respectively.
Each grain species initially moves coupling with the gas with the 
velocity of $\sim$1300, $\sim$900, and $\sim$400 km s$^{-1}$ for C, 
Mg$_2$SiO$_4$, and Fe grains, respectively, and collides with the
reverse shock at 3650 yr for C grains formed in the outermost He core, 
6300 yr for Mg$_2$SiO$_4$ grains in the oxygen-rich layer, and 13000 yr
for Fe grains condensed in the innermost He core.

The collision time of the reverse shocks with dust grains depends on the 
initial velocity and position of dust, the thickness of the hydrogen 
envelope, and the density in the ISM.
Figure 4 gives the collision time $t_{\rm coll}$ of the reverse shock
with the He core for different progenitor mass and gas density in the
ISM.
We can see that the collision time is shorter for PISNe than SNe II, in 
spite of the fact that PISNe have the thicker hydrogen envelopes.
This reason is that the gas velocities ($\sim$2000--3000 km s$^{-1}$) at 
the outermost He core for PISNe with the explosion energies higher than
10$^{52}$ erg are a few times higher than those for SNe II.
For $n_{\rm H,0} =$ 0.1--10 cm$^{-3}$, the collision times are 
$\simeq$10$^3$--10$^4$ yr and decrease with increasing the ambient gas 
density.
Note that this result is not consistent with the observations of the Cas
A SNR, where thermal emission from dust heated by the reverse shock has 
already detected at $\simeq$330 yr after the explosion (Ennis et al. 
2006 and references therein).
This is because the progenitor of the Cas A is believed to have lost the 
considerable hydrogen envelope during their evolution (Young et al. 
2006), in contrast to the Population III SNe considered in this paper.
We shall discuss the effect of the hydrogen envelope on the evolution of
dust in the SNR in \S~4.

The fates of dust grains within SNRs heavily depend on their initial
sizes $a_{\rm ini}$ as well as the chemical composition reflecting the
difference in the sputtering yield and bulk density, as shown in Figure
3.
For the standard model, the relatively small grains with $a_{\rm
ini}=0.01$ $\micron$ are efficiently decelerated due to the gas drag and 
are fully trapped into the hot plasma to be completely destroyed by the
thermal sputtering.
Note that the gas drag on grains with small radii is more efficient than 
that on larger grains, because the deceleration rate of grain is
inversely proportional to its size (NKH06).
Actually, the grains with the initial sizes less than 0.05 $\micron$
are trapped into the hot gas between the forward and reverse shocks and 
continue to be eroded by the thermal sputtering even at over $10^5$ yr 
until they are completely destroyed.
Larger grains of $a_{\rm ini} = 0.1$ $\micron$ undergo the kinetic
and thermal sputtering while streaming in the hot gas. 
Thanks to the high bulk density, Fe grains with $a_{\rm ini}=0.1$
$\micron$ are injected into the ISM, reducing the size by 52 \%. 
On the other hand, C and Mg$_2$SiO$_4$ grains with $a_{\rm ini}=0.1$ 
$\micron$ are trapped and eroded by the thermal sputtering in the denser 
region behind the forward shock, and their surface layers whose 
thicknesses are 43 and 69 \% of their initial sizes are eroded, 
respectively, until $2 \times 10^5$ yr when the SNR enters into the 
radiative phase and the dense SN shell is formed behind the forward
shock. 
These dust grains remain in the dense shell without further processing 
because the gas cools down quickly below $10^6$ K.
Thus, the decrease of sizes of these grains is truncated at a given size,
and the erosion of dust with $a_{\rm ini} =$ 0.05--0.2 $\micron$ results 
in the final size of 0.001--0.1 $\micron$, depending on their initial 
sizes.
For 1 $\micron$-sized C, Mg$_2$SiO$_4$, and Fe grains, the kinetic 
sputtering reduces their sizes by 0.7, 6, and 8 \%, respectively.
Note that large grains with $a_{\rm ini} \ga 0.2$ $\micron$ can go across 
even the forward shock and be injected into the ISM, because the 
deceleration due to the gas drag is very inefficient.

We should mention here that the degree of the erosion of dust is a 
complex function of the initial position and initial size of the dust as 
well as the sputtering yield.
Fe grains formed in the innermost He core and travelling through the
oxygen-rich gas undergo the efficient erosion, since the sputtering
yield by an oxygen ion is $\sim$50 times larger than that by a proton.
In contrast, the degree of the erosion of C grains in the outermost He 
core is relatively small because they can quickly escape from the He 
core after the collision with the reverse shock and also have the 
sputtering yield lower than those of other dust species. 
Furthermore, without being decelerated efficiently, the larger grains
with the high velocities relative to gas are more efficiently eroded by 
sputtering in the relatively dense region near the forward shock front.
However, the grains with the sizes larger than a given size can evade
the erosion in the shocked hot gas within SNRs and are expelled into the 
ISM.

Although the processing of dust in the ISM is not the main subject of 
this paper, here we shall simply show the processing of dust injected
into the ISM as a consequence of dust evolution in SNRs.
Dust grains injected into the ISM can be consumed through the kinetic 
sputtering because of the difference in velocity between the ambient
cool gas and the dust grains, while the grains can be also decelerated
by the direct collisions with the gas.
In this case, the ratio of the final radius $a_{\rm fin}$ to the initial 
size $a_{\rm ini}$ of dust is dependent only on the escape velocity
$w_0$ defined as the velocity with which dust is injected into the ISM 
passing through the forward shock, and is given by
\begin{eqnarray}
\frac{a_{\rm fin}}{a_{\rm ini}} = \exp \left[ - \left( 
\frac{2 m_{\rm sp}}{3 \mu_g} \right) \int^{w_0}_{w_{\rm f}} 
\sum_i A_i Y^0_i(w) \frac{dw}{w} \right], 
\end{eqnarray}
where $m_{\rm sp}$ is the average atomic mass of the elements sputtered 
from the grain, $\mu_g$ is the mean molecular weight of the gas, and 
$Y^0_i(w)$ is the sputtering yield at normal incidence by gas species
$i$ whose number abundance is $A_i$.

Figure 5 shows the results calculated by Equation (1) for the primordial
gas composition as a function of the escape velocity.
In the calculations, the final relative velocity $w_{\rm fin}$ is taken 
as 10 km s$^{-1}$, which is small enough for dust grains not to be
eroded by the kinetic sputtering.
Also we assume that the escape velocity equals to the initial velocity 
inside the He core, since very large grains are ejected to the ISM with
high velocities, not being decelerated efficiently.
Hence, the calculated $a_{\rm fin}$ gives the lower limit of the final 
size realized in the ISM.
The final size acquired by each dust species is different, depending on 
the sputtering yield $Y^0_i$ and the average atomic mass $m_{\rm sp}$, 
and decreases with increasing the escape velocity.
The ratio of the final size to the initial size is $\simeq$0.8 for C 
grains with $w_0 \sim 1300$ km s$^{-1}$, $\simeq$0.5 for Mg$_2$SiO$_4$ 
grains with $w_0 \sim 900$ km s$^{-1}$, and $\simeq$0.5 for Fe grains
with $w_0 \sim 400$ km s$^{-1}$.
Thus, the sizes of very large grains supplied from SNe are decreased to 
0.5--0.8 times those at the time of the ejection, but are not completely 
destroyed in the ISM.
For PISNe with the explosion energies higher than $10^{52}$ erg, the
sizes of Fe grains with the initial velocities of $\simeq$1000 km
s$^{-1}$ decrease by 70 \% in the ISM, although the sizes of C and 
Mg$_2$SiO$_4$ grains whose initial velocities are in the range of 
2000--3000 km s$^{-1}$ are not significantly different from those 
calculated for the standard model. 
Note that dust grains in the ISM are also processed by the high-velocity 
interstellar shocks driven by the ambient SNe (NKH06), which is the
major mechanism of the dust destruction in the ISM.

\subsection{\textit{Efficiency of Dust Destruction}}

The results described in \S~3.2 imply that the size distribution of
survived dust is greatly deficient in small-sized grains, compared with 
that at the time of dust formation.
In Figure 6, we present the initial size distribution at the time of
dust formation (Fig. 6a) and the size distribution of dust at the 
truncation time (Fig. 6b) for the standard model.
The comparison of these figures clearly indicates that grains with
the radii below a few tens \AA~are missing for almost all dust species.
In particular, Al$_2$O$_3$ grains are completely destroyed because their
initial sizes are smaller than 0.02 $\micron$.
We can also see that the sizes of the relatively large grains are
shifted to small sizes due to the erosion by sputtering.
It should be pointed out here that the size distribution of survived
dust is different from that by Bianchi \& Schneider (2007): 
They have found that the final size distribution show a flattening
towards smaller sizes without any abrupt truncation.
The main reason of this difference is considered as follows.
Bianchi \& Schneider (2007) assumed that newly formed grains remain 
confined and trapped at their initial positions in the ejecta.
Therefore, dust grains remaining at their initial positions cease to 
be eroded by sputtering on the timescale of 4--8$\times 10^4$ yr due to
the decrease of the gas density caused by expansion.
As shown in \S~3.2, our results show that even small grains of $a_{\rm
ini} \le 0.05$ $\micron$ penetrate into the hot plasma between the
forward and reverse shocks, and are trapped and completely destroyed in
the hot gas with the relatively high density.  
On the other hand, the erosion of grains with the radii of 0.05--0.2 
$\micron$ produces the grains with the final sizes of 0.001--0.1 
$\micron$ remaining in the dense shell.

The critical size below which dust is destroyed is sensitive to the gas 
density in the ambient medium.
Although the critical size weakly depends on the dust species and their 
initial positions, we can roughly estimate the average critical size for 
different gas density in the ISM.
For $n_{\rm H,0}=0.1$ cm$^{-3}$, the upper limit of the initial size of 
dust completely destroyed is $\sim$0.01 $\micron$, and the lower limit
of that ejected to the ISM is $\sim$0.03 $\micron$.
For $n_{\rm H,0}=10$ cm$^{-3}$, the grains with $a_{\rm ini} \la 0.2$ 
$\micron$ are destroyed, and the grains with $a_{\rm ini} \ga 0.5$ 
$\micron$ are injected in the ISM.
Note that the above critical sizes are true for the grain species except 
for C grains.
The critical size of C grain is 0.006, 0.02, and 0.07 $\micron$ for 
$n_{\rm H,0}=$0.1, 1, and 10 cm$^{-3}$, respectively, and is a few times 
smaller than other dust species, because C grains are located at the 
outermost He core and also have the lower erosion rate by sputtering.
Note that the initial sizes of Fe grains injected into the ISM are
smaller than those of others by a factor of $\sim$2 because of the high 
bulk density, and are $\ga$0.02, 0.1, and 0.25 $\micron$ for 
$n_{\rm H,0}=$0.1, 1, and 10 cm$^{-3}$, respectively.

On the other hand, the critical size for each grain species is almost 
independent of the progenitor mass as long as the explosion energy of SN 
is the same, because the time evolution of temperature and density of
gas within SNRs is similar.
For PISNe with the explosion energies higher than $10^{52}$ erg and more 
massive hydrogen envelopes, the critical size increases approximately by 
a factor of three, compared with that for SNe II.
However, the initial size of dust ejected to the ISM is less than 2 
times that for SNe II, because the high initial velocity inside the He 
core makes the dust grains easily escape from SNRs.
These results lead to the conclusion that the large-sized grains
dominate the mass of dust injected from SNe into the ISM.

Tables 1 and 2 summarize the mass fraction of dust destroyed
$\epsilon_{\rm dest}$, piled up in the dense shell $\epsilon_{\rm
shell}$, and ejected to the ISM $\epsilon_{\rm eject}$ for the unmixed 
and mixed grain models, respectively.
In Figures 7a and 7b, we also present the total mass of survived dust 
for both grain models, respectively, along with their initial total
mass.
We can first see that the mass of dust destroyed increases with 
increasing the surrounding gas density.
In particular, for $n_{\rm H, 0}=10$ cm$^{-3}$, all or almost all 
($\ga$85 \%) of dust grains formed in the ejecta are destroyed, and the 
mass of survived dust is less than $0.1$ $M_\odot$ for all models 
considered in this paper.
This reason is as follows; the erosion of dust takes place in the hot
gas between the forward and reverse shocks, whose density increases with 
the ambient gas density. Therefore, the higher ISM gas density leads to 
the efficient erosion and deceleration of dust through the more frequent 
collisions with the hot gas.
Next, we find that the mass fraction of dust destroyed is generally 
higher for the mixed grain model than the unmixed grain model; 
for SNe II with $n_{\rm H, 0}=$ 1 (0.1) cm$^{-3}$, $\epsilon_{\rm dest}
=$ 0.5--0.8 (0.2--0.4) for the unmixed grain model, while 
$\epsilon_{\rm dest} =$ 0.9--0.99 (0.57--0.78) for the mixed grain
model.
This reflects the fact that the mixed grain model lacks the grains 
larger than 0.05 (0.01) $\micron$ in comparison with the unmixed grain 
model.
Finally, it can be seen that dust grains formed in PISNe are more
efficiently destroyed than those in SNe II, since the newly formed dust 
grains are dominated by the small-sized grains and the critical size is 
much larger.
Thus, for $n_{\rm H, 0}=1$ cm$^{-3}$, the mass of dust survived in 
PISNe is 0.1--1 $M_\odot$ for both grain models and is almost 
the same as that (0.07--0.5 $M_\odot$) in the SNe II, though the mass of 
dust formed in PISNe is a few tens times higher than that in SNe II.
Note that the dust destruction by reverse shocks is more effective 
than that by high-velocity interstellar shocks for the dust grains with 
the same size distribution (NKH06).

The destruction efficiency defined as the ratio of the mass of dust
destroyed to the initial dust mass is given in Tables 3 and 4 for each
dust species in the unmixed and mixed grain model, respectively.
As mentioned above, the efficiency of dust destruction is greatly
influenced by what fraction of the initial dust mass is occupied by the 
sizes larger than the critical size.
Thus, the destruction efficiency for a given dust species is very 
sensitive to the initial size distribution.
Note that it is difficult to find the clear dependence of the
destruction efficiency of each dust species on the progenitor mass
because the size distribution of dust at the time of dust formation is 
different from model to model.
However, for example, the efficiency of destruction of Al$_2$O$_3$ grain 
with $a_{\rm ini} \la 0.02$ $\micron$ is $\sim$1 for almost all cases 
considered in this paper.
For the unmixed grain model, FeS and MgSiO$_3$ grains are also 
predominantly destroyed.
On the other hand, Fe, Si, and C grains for which most of the mass is 
locked into the grains larger than 0.05 $\micron$ have the relatively 
small destruction efficiencies.
For the mixed grain model, SiO$_2$ grains are the main dust species that
are left in SNRs and/or are injected into the ISM.
Therefore, we conclude that the chemical composition, size distribution, 
and amount of dust grains supplied from SNe to the ISM are quite
different from those at the time of dust formation.

Here we shall discuss the dependence of the metallicity on the present 
results of calculations. 
First, it should be pointed out that the species of dust formed in the 
ejecta of SNe II and their size distributions are not sensitive to the 
metallicity of progenitor stars (Todini \& Ferrara 2001; Nozawa 2003). 
Also the cooling function of gas for $Z \le 10^{-3}$ $Z_{\odot}$ is 
independent of the metallicity (Sutherland \& Dopita 1993). 
Thus, the results of calculations presented in this paper can be
directly applied to the evolution of dust in the ejecta of SNe II
expanding into the ambient medium whose metallicity is less than 
10$^{-3}$ $Z_{\odot}$.
The increase of the metallicity in the ambient medium to the solar value 
greatly enhances the cooling of gas behind the forward shock and causes 
the time of the transition from nonradiative phase to radiative phase to 
be reduced to less than half of that for the zero-metallicity case. 
Nevertheless, the destruction efficiency of each grain species for 
$Z = Z_\odot$ decreases at most by 15 \% of that for $Z=0$. 
Therefore, the results of present study could be useful for evaluating 
the dust evolution in SNRs generated from SNe II, regardless of the 
initial metallicity of progenitor stars and ambient medium.
However, as is demonstrated in the next section, the thickness of
hydrogen envelope strongly affects the motion and destruction of dust 
within SNRs.

\section{\textit{THE EFFECT OF THE HYDROGEN ENVELOPE ON THE EVOLUTION OF
 DUST IN SNR}}

As mentioned in \S~3.2, for the SNe with the thick hydrogen envelopes, 
it takes at least more than 1000 yr for the reverse shocks to collide 
with the dust condensed inside the He core.
On the other hand, the infrared observations of the Cas A SNR (Dwek et
al. 1987; Lagage et al. 1996; Arendt et al. 1999; Douvion et al. 2001; 
Hines et al. 2004; Ennis et al. 2006) have revealed the thermal emission 
from warm dust formed in the ejecta at $\simeq$330 yr after the
explosion.
This difference of the time at which dust grains inside the He core are
swept up by the reverse shocks is attribute to the difference in the
thickness of the hydrogen envelope of SNe.
Therefore, in this section, we investigate the effect of the hydrogen 
envelope on the evolution of dust in the SNR.

In order to illustrate the evolution of dust in the SNR generated from
the SN whose hydrogen envelope is very thin, we adopt the model of
ejecta with $M_{\rm pr} = 20$ $M_\odot$ from Umeda \& Nomoto (2002) and 
modify it by artificially reducing the mass of the hydrogen envelope
from 13 $M_\odot$ to 0.7 $M_\odot$ at the time of explosion.
Keeping the structure of the density, we simply scale up the gas
velocity so that the explosion energy can equal to 10$^{51}$ erg.
We also assume the dust inside the He core is the same as that formed in
primordial SN II with $M_{\rm pr} = 20$ $M_\odot$.

Figure 8 illustrates the trajectories (Fig. 8a) and time evolutions of
sizes (Fig. 8b) of C, Mg$_2$SiO$_4$, and Fe grains with $a_{\rm ini} =$ 
0.01 and 0.1 $\micron$ in the SNR calculated for $n_{\rm H,0}=$ 1 
cm$^{-3}$.
As can be expected, because the hydrogen envelope is very thin and the 
initial velocities of dust grains are very high (2000--5000 km s$^{-1}$
depending on their initial positions), the dust grains inside the He
core collide with the reverse shock at much earlier time;
120 yr for C grains, 570 yr for Mg$_2$SiO$_4$ grains, and 1300 yr for Fe 
grains.
Contrary to SNe II with the thick hydrogen envelopes, even the grains 
with $a_{\rm ini} = 0.1$ $\micron$ can be ejected to the ISM without 
decreasing their sizes significantly; 
C and Mg$_2$SiO$_4$ grains with $a_{\rm ini} = 0.1$ $\micron$ reduce
their sizes only by less than 6 \%, while the sizes of Fe grains with 
$a_{\rm ini} =0.1$ $\micron$ decrease by 18 \% because the time staying
in the hot gas is long, compared with C and Mg$_2$SiO$_4$ grains.
The grains with $a_{\rm ini} = 0.01$ $\micron$ are trapped into the hot 
gas and are completely destroyed.
In this case, the critical size below which dust is completely destroyed 
is $\sim$0.04 $\micron$, and the lower limit of the initial size of dust 
supplied to the ISM is $\sim$0.04 $\micron$ which is five times smaller 
than that for a SN II.
Therefore, the thin hydrogen envelope as well as the high initial
velocity of dust causes almost all survived dust grains to be injected 
into the ISM without being trapped into the gas within the SNR.
The mass fraction of dust destroyed is 0.38, which is lower than 0.75
for the standard model, and thus more dust grains are supplied to the
ISM.
The result of calculation shows that the fates of dust grains formed in
the ejecta strongly depend on the thickness of the hydrogen envelope.

However, we adopt the model of dust formed in a SN II and also consider
the uniform ambient medium.
The significant mass loss of massive stars during their evolution
results in the circumstellar medium that is not uniform and homogeneous.
In addition, the time evolution of the temperature and density of gas in 
the ejecta of SNe without the hydrogen envelopes such as Type Ib/c SNe
is expected to be different from that in SNe II, which influences the
dust formation in the ejecta.
Therefore, we must clarify the size distribution and amount of dust 
grains formed in the ejecta to investigate the dependence of the 
evolution of dust on the type of SNe. 
This subject will be reported in the forthcoming paper.

\section{\textit{ELEMENTAL ABUNDANCES OF THE SECOND-GENERATION STARS}}

Some extremely metal-poor (EMP) stars with [Fe/H] $\la -2.5$ discovered
in the Galactic halo have peculiar abundance patterns showing the modest 
or large (1--100 times) enhancements of C, N, O, Mg, and Si relative to 
solar (Venkatesan et al. 2006 and references therein). 
In particular, two hyper metal-poor (HMP) stars with [Fe/H] $\la -5$, HE 
0107-5240 ([Fe/H] $=-5.2$, Christlieb et al. 2002) and HE 1327-2326 
([Fe/H] $=-5.4$, Frebel et al. 2005) show extreme ($\ga$10$^2$ times) 
overabundances of C, N, and O relative to iron.
Because the elemental composition of HMP stars expected to be the very 
early generation may strongly reflect the nucleosynthesis in Population 
III stars, several scenarios have been proposed to explain the origin of 
elemental abundances in low-mass HMP stars; 
the mixing fall-back in a core-collapse SN (Umeda \& Nomoto 2005;
Iwamoto et al. 2005), the nucleosynthesis and mass transfer in a 
first-generation binary star (Suda et al. 2004; Komiya et al. 2006), 
the pollution by the gas with heavy elements in the ISM (Shigeyama et
al. 2003), and the combinations of these scenarios described above 
(Christlieb et al. 2004).

Recently, Venkatesan et al. (2006) have proposed that EMP stars are the 
second-generation stars formed in the dense shell of primordial SNRs and 
their peculiar abundances can be reproduced by the segregated transport
of newly formed dust decoupled from the metal-rich gas in SNe.
Adopting the dust models by Todini \& Ferrara (2001) and Schneider et 
al. (2004), they calculated the sputtering and the transport of dust
driven by the UV radiation field from the stellar cluster within a SNR. 
They have conclude that the progenitor mass range of 10--150 $M_\odot$ 
qualitatively explains the enhancement of the elements composing dust 
grains (C, O, Mg, and Si) in EMP stars, though they did not 
compare their results with the abundance data on these metal-poor stars.

The results of the transportation and destruction of dust given in
\S~3.2 show that all dust grains remaining in primordial SNRs without 
being completely destroyed by sputtering can be accumulated in the dense
SN shell at $\sim$10$^5$--10$^6$ yr after the explosion.
Thus, it is considered that the formation of the second-generation stars
with solar mass scales could be possible in the shell contaminated with
dust grains, as is investigated by Schneider et al. (2006).
In this case, we can expect that the elemental compositions of dust
grains piled up into the dense shell reflect the metal abundance
patterns of the second-generation stars.

The iron-bearing dust species including the most important element Fe 
are Fe and FeS grains for the unmixed grain model, and Fe$_3$O$_4$ grain 
for the mixed grain model.
Note that C grains cannot be formed in the mixed ejecta 
because the ejecta is oxygen-rich, and
Fe$_3$O$_4$ grains with $a_{\rm ini} \la 0.05$ $\micron$ are dominantly 
destroyed in the postshock flow and are rarely accumulated in the SN 
shell for $n_{\rm H,0}=$ 1 and 10 cm$^{-3}$.
Hence, we show only the results calculated for the unmixed grain model.
In addition, Fe and FeS grains are not significantly piled up in the 
shell of PISN remnants, since most of them are completely destroyed or 
are injected into the ISM.
Thus, we focus on the results for Type II SNRs.

The abundances of C, O, Mg, and Si relative to Fe in the dense shell 
are summarized in Tables 5, where we neglect the contribution of metal 
atoms sputtered from the grains crossing the dense shell 
because the largest-sized grains are only eroded by less than 1\% of their
radii.
The hydrogen mass in the dense SN shell $M^{\rm H}_{\rm shell}$ at the 
truncation time is given in Table 1.
It can be seen from Table 5 that all models considered here predict the 
value of [Fe/H] less than $-4$, and most of them exhibit the value of 
[Fe/H] ranging from $-6$ to $-5$.
Thus, two HMP stars are considered to be formed in the dense shell of 
primordial SNRs.
Among the 9 models with $-6 \le$ [Fe/H] $\le -5$, the abundances of Mg
and Si are in the range of $-1.2 \le$ [Mg/Fe] $\le 1.2$ and 
$-0.6 \le$ [Si/Fe] $\le 2.7$, respectively, and 8 models can produce
the overabundances of Mg and/or Si.
This result suggests that the dust formed in the unmixed ejecta of SNe
II can be responsible for the peculiar abundance patterns of Mg and Si
in HMP stars.
Although a few models (20 $M_\odot$ and 30 $M_\odot$ for $n_{\rm
H,0} =$ 1 cm$^{-3}$ and 25 $M_\odot$ for $n_{\rm H,0} =$ 0.1 cm$^{-3}$)
result in the overabundances of C and O as well as Mg and Si, [C/Fe] and
[O/Fe] are limited to $<$1.6 and $<$0.6, respectively.

For comparison, we apply the present calculations to the models of dust 
formation in the SN II with $M_{\rm pr} =$ 22 $M_\odot$ by Todini \& 
Ferrara (2001) and in PISN with $M_{\rm pr} =$ 195 $M_\odot$ by
Schneider et al. (2004).
Because the sizes of the dust grains calculated by them are considerably 
small ($\la$0.04 $\micron$) except for C grain, the grain species that can 
survive the destruction through the collisions with the reverse shocks
is only C grain for the ambient gas density of $n_{\rm H,0}=$ 0.1--10 
cm$^{-3}$.
Thus, their dust models cannot explain the abundance patterns of the
metals except for C observed in HMP and EMP stars.

\section{SUMMARY}

We investigate the evolution of dust formed at Population III SN 
explosions through the collision with the reverse shocks within SNRs.
We adopt the models of dust grains obtained from the calculation of dust 
formation in Population III SNe by Nozawa et al. (2003) and take into 
account their spatial distribution and size distribution as the initial
condition.
The calculations carefully treat the dynamics, erosion, and heating of 
dust grains, and the evolution of temperature and density of gas in 
spherically symmetric shocks is solved as a function of time.
We also discuss the effect of the hydrogen envelope on the evolution 
of dust in the SNR.
Furthermore, from the analysis of the transport and destruction of dust 
within SNRs, we investigate the abundances of elements related to dust
in the second-generation stars formed in the dense shell of primordial
SNRs.

The our main results are summarized as follows.

1. The time that the reverse shock encounters the dust that condensed 
 inside the He core depends on the thickness of the H-envelope that was 
 retained by the progenitors of the Population III SNe. 
 If the progenitor stars did not undergo significant mass loss, the
 reverse shock will encounter the dust $\simeq$$10^3$--10$^4$ yr after
 the explosion, depending on the density of gas in the ISM.
 
 2. Once dust grains inside the He core collide with the reverse shocks, 
 they will follow different trajectories depending on their initial
 sizes, resulting in the differential transport and destruction of dust in 
 SNRs.
 For Type II SNRs expanding into the ISM with the density of 
 $n_{\rm H,0}=1$ cm$^{-3}$, small grains with $a_{\rm ini} \la 0.05$ 
 $\micron$ are rapidly trapped into the postshock flow and are 
 completely destroyed by sputtering. 
 Grains with $a_{\rm ini}=$ 0.05--0.2 $\micron$ are trapped and remain 
 in the dense shell behind the forward shock.
 Very large grains with $a_{\rm ini} \ga 0.2$ $\micron$ are ejected to
 the ISM through the forward shock without significantly decreasing
 their sizes.

 3. The critical size below which dust is completely destroyed in SNRs
 is sensitive to the gas density in the ambient medium, and spans the
 range of 0.01--0.2 $\micron$ for $n_{\rm H, 0} =$ 0.1--10 cm$^{-3}$. 
 The resulting size distribution of survived dust is greatly deficient
 in small-sized grains, compared with that at the time of dust formation,
 although the erosion of large grains produces the smaller-sized grains.
 Thus, the mass of dust injected from SNe into the ISM is dominated by
 the large grains.

 4. The total mass fraction of dust destroyed in SNRs ranges from 0.2 to
 1 and increases with increasing the ambient gas density and explosion 
 energy of SNe.
 The destruction efficiency of each dust species is very sensitive to
 the initial size distribution, and the dust species whose mass is 
 predominantly occupied by the sizes larger than the critical size can 
 survive.
 Therefore, the chemical composition, size distribution, and amount of 
 dust grains supplied from SNe to the ISM are quite different from those 
 at the time of dust formation.
 
 5. The results for the evolution of dust in SNRs presented in this
 paper can be directly applied for the initial metallicity of progenitor 
 stars and ambient medium less than $Z \le 10^{-3}$ $Z_\odot$, and could 
 be useful for evalulating the evolution of dust in the Galactic SNRs 
 generated from SNe II.
 
 6. The fates of dust grains formed in the ejecta strongly depend on the 
 thickness of the hydrogen envelope.
 For the SNR generated from the SN with the very thin hydrogen envelope, 
 the collision time of the reverse shock with dust grains inside the He
 core is much earlier. 
 As long as the model of dust for SNe II is employed for the
 calculation, the mass of dust supplied to the ISM is larger than SNe II 
 with the thick hydrogen envelopes.
 
 7. If the elemental compositions of dust grains piled up in the SN
 shell reflect the metal abundance patterns of the second-generation
 stars formed in the dense shell of primordial SNRs, the dust formed in
 the unmixed ejecta of SNe II can be responsible for the peculiar
 abundance patterns of Mg and Si in HMP stars.
 However, another scenario could be necessary to produce the large 
 overabundances of C and O observed in HMP stars.

\acknowledgments

The authors are grateful to the anonymous referee for critical comments
that are useful for improving the manuscript.
This work has been supported in part by a Grant-in-Aid for Scientific
Research from the Japan Society for the Promotion of Sciences
(16340051 and 18104003).

\clearpage

\begin{figure}
\epsscale{0.7}
\plotone{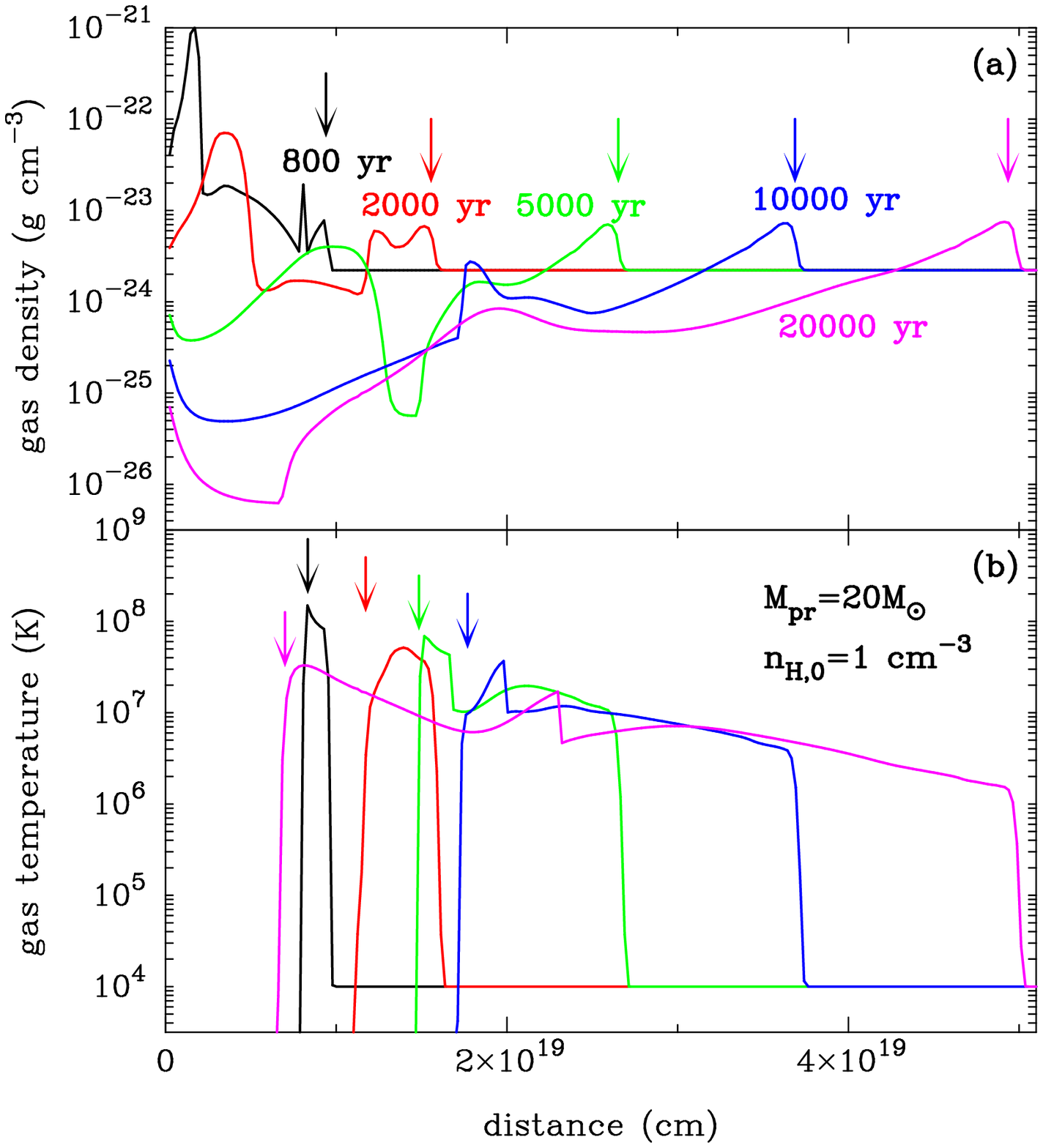}
\caption{
 The structures of (a) the density and (b) the temperature of the gas at 
 given times within the SNR generated from the explsoion of star with 
 $M_{\rm pr} = 20$ $M_\odot$ and expanding into the ISM with 
 $n_{\rm H,0}=1$ cm$^{-3}$ (the standard model).
 The positions of the forward and reverse shocks are indicated by the 
 downward arrows in (a) and (b), respectively. 
\textit{See the electronic edition of the Journal for a color version 
of this figure.}\label{fig1}}
\end{figure}

\clearpage

\begin{figure}
\epsscale{0.7}
\plotone{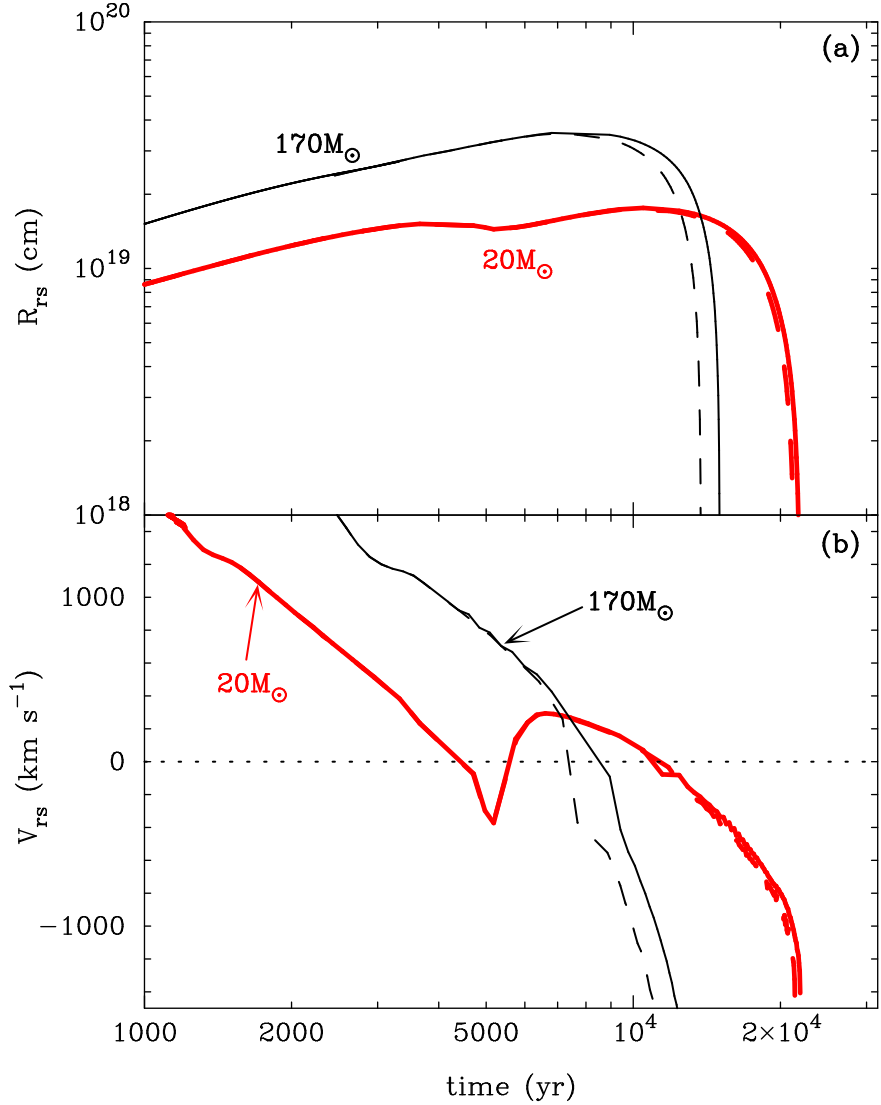}
\caption{
 The time evolution of (a) the position $R_{\rm rs}$ and (b) the
 velocity $V_{\rm rs}$ of the reverse shock.
 The thick solid curves represents the results for the standard model
 with $M_{\rm pr} = 20$ $M_\odot$ and $n_{\rm H,0}=1$ cm$^{-3}$, and the 
 thin solid curves for the model with $M_{\rm pr} = 170$ $M_\odot$ and 
 $n_{\rm H,0}=1$ cm$^{-3}$.
 The results that do not include the cooling by thermal emission from
 dust are also shown by the thick and thin dashed curves for 
 $M_{\rm pr} = 20$ $M_\odot$ and 170 $M_\odot$, 
 respectively.
\textit{See the electronic edition of the Journal for a color version 
of this figure.}\label{fig2}}
\end{figure}

\clearpage

\begin{figure}
\epsscale{0.7}
\plotone{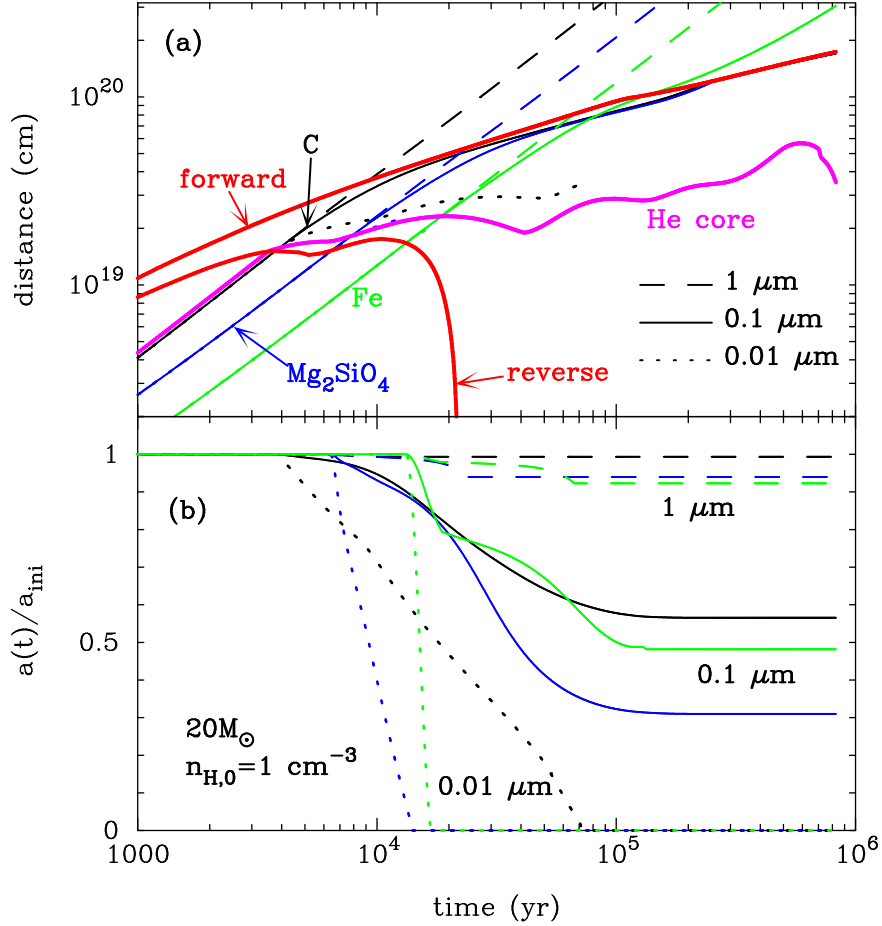}
\caption{
 The time evolutions of (a) the positions and (b) the ratios of size to 
 the initial size of dust grains within the SNR for the standard model.
 The positions of the forward and reverse shocks are depicted by the 
 thick solid curves in (a), along with the position of the surface of 
 the He core.
 Among nine dust species for the unmixed grain model, are shown C, 
 Mg$_2$SiO$_4$, and Fe grains with the initial sizes of 0.01 $\micron$ 
 (\textit{dotted lines}), 0.1 $\micron$ (\textit{solid lines}), and 1
 $\micron$ (\textit{dashed lines}), respectively.
\textit{See the electronic edition of the Journal for a color version 
of this figure.}\label{fig2}}
\end{figure}

\clearpage

\begin{figure}
\epsscale{0.7}
\plotone{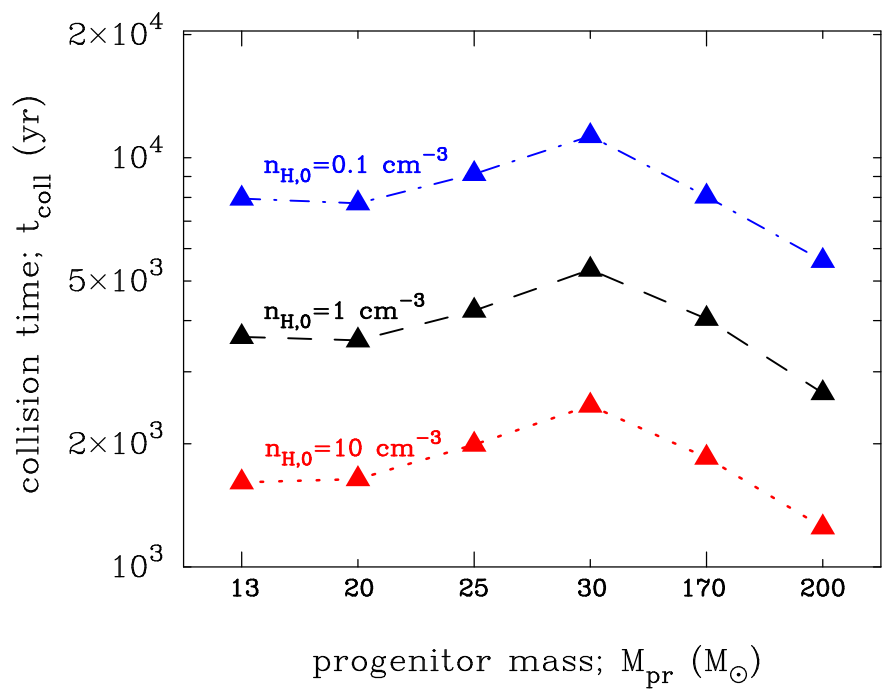}
\caption{
 The collision time $t_{\rm coll}$ of the reverse shocks with the He
 core vs. the progenitor stellar mass for different gas density in the
 ISM.
 The results for $n_{\rm H,0}=$ 0.1, 1, and 10 cm$^{-3}$ are connected
 by the dot-dashed, dashed, and dotted lines, respectively.
\textit{See the electronic edition of the Journal for a color version 
of this figure.}\label{fig4}}
\end{figure}

\clearpage

\begin{figure}
\epsscale{0.7}
\plotone{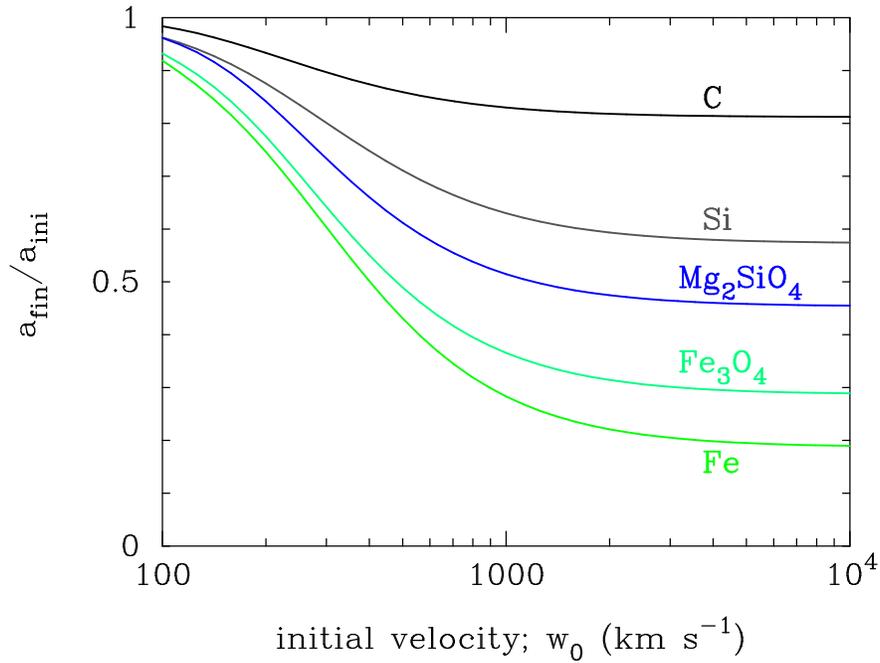}
\caption{
 The ratio of the final size $a_{\rm fin}$ to the initial size 
 $a_{\rm ini}$ of dust eroded by the kinetic sputtering in the ISM 
 with the primordial gas composition as a function of the escape 
 velocity $w_0$, where the final relative velocity $w_{\rm fin}$ is
 taken as 10 km s$^{-1}$.
 The results are shown for C, Si, Mg$_2$SiO$_4$, Fe$_3$O$_4$, and Fe
 grains.
\textit{See the electronic edition of the Journal for a color version 
of this figure.}\label{fig7}}
\end{figure}

\clearpage

\begin{figure}
\epsscale{0.7}
\plotone{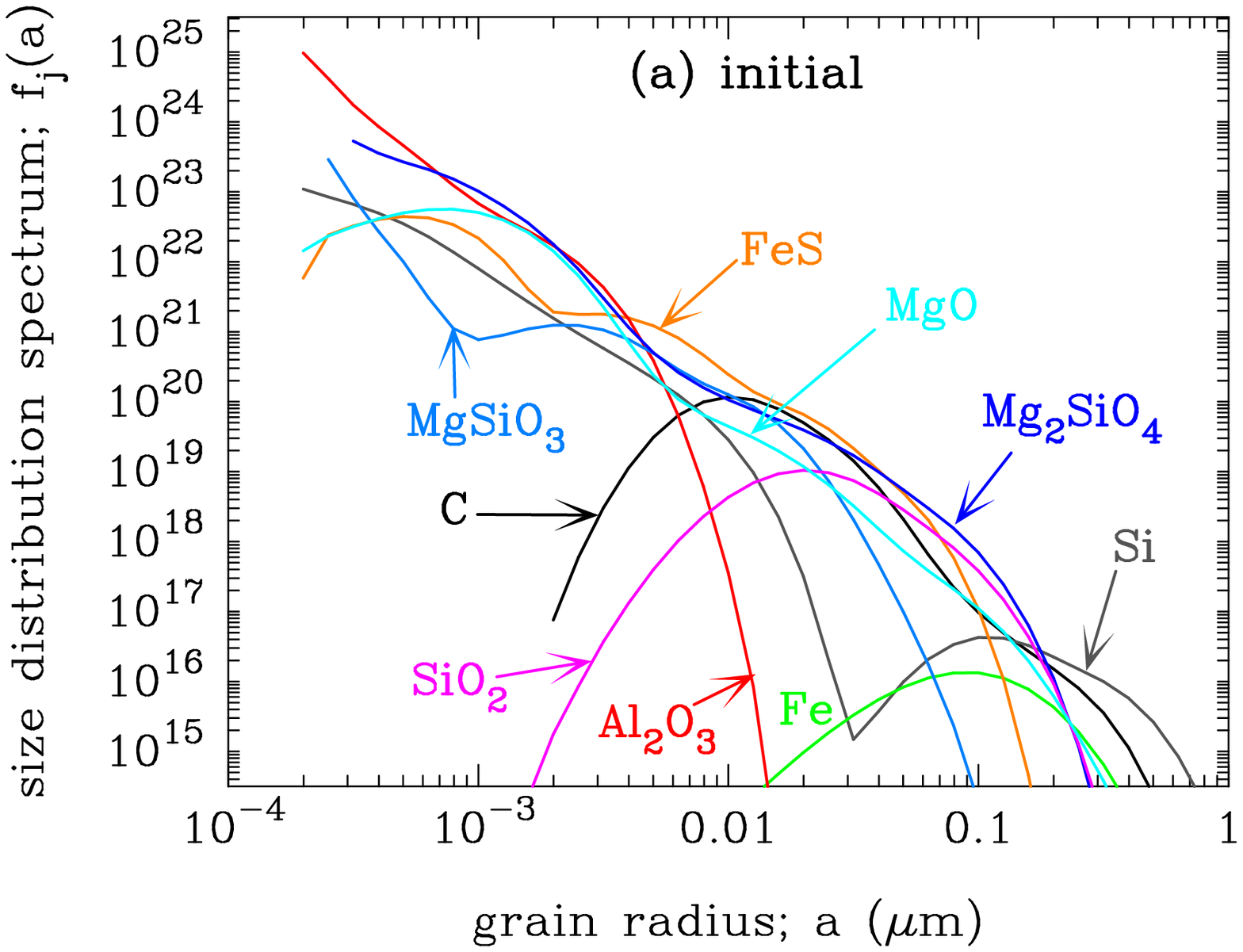}
\plotone{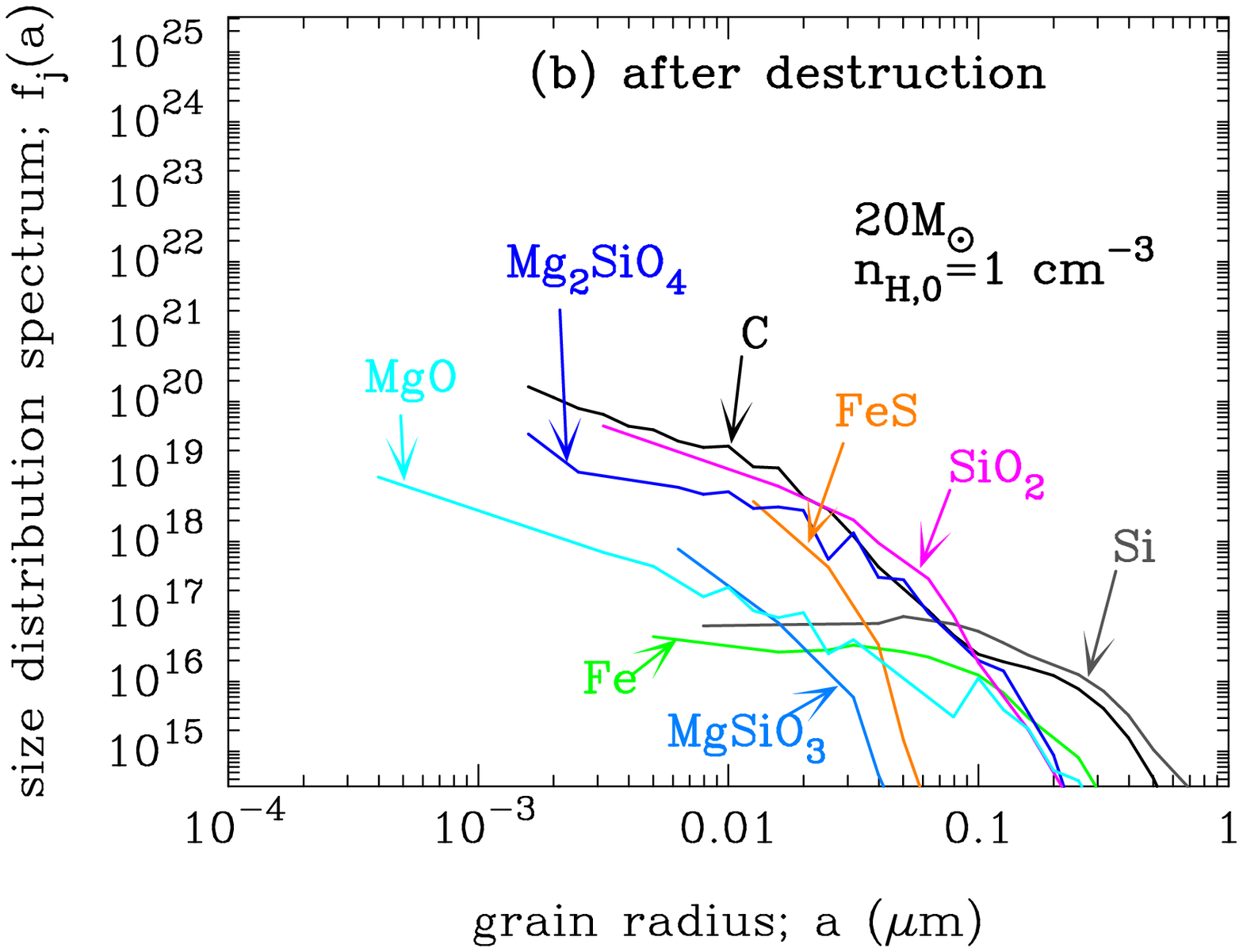}
\caption{ 
 The size distribution of each dust species for the standard model;
 (a) for the initial size distribution before destruction and 
 (b) for the resulting size distribution of survived dust after
 destruction.
\textit{See the electronic edition of the Journal for a color version 
of this figure.}\label{fig6}}
\end{figure}

\clearpage

\begin{figure}
\epsscale{0.7}
\plotone{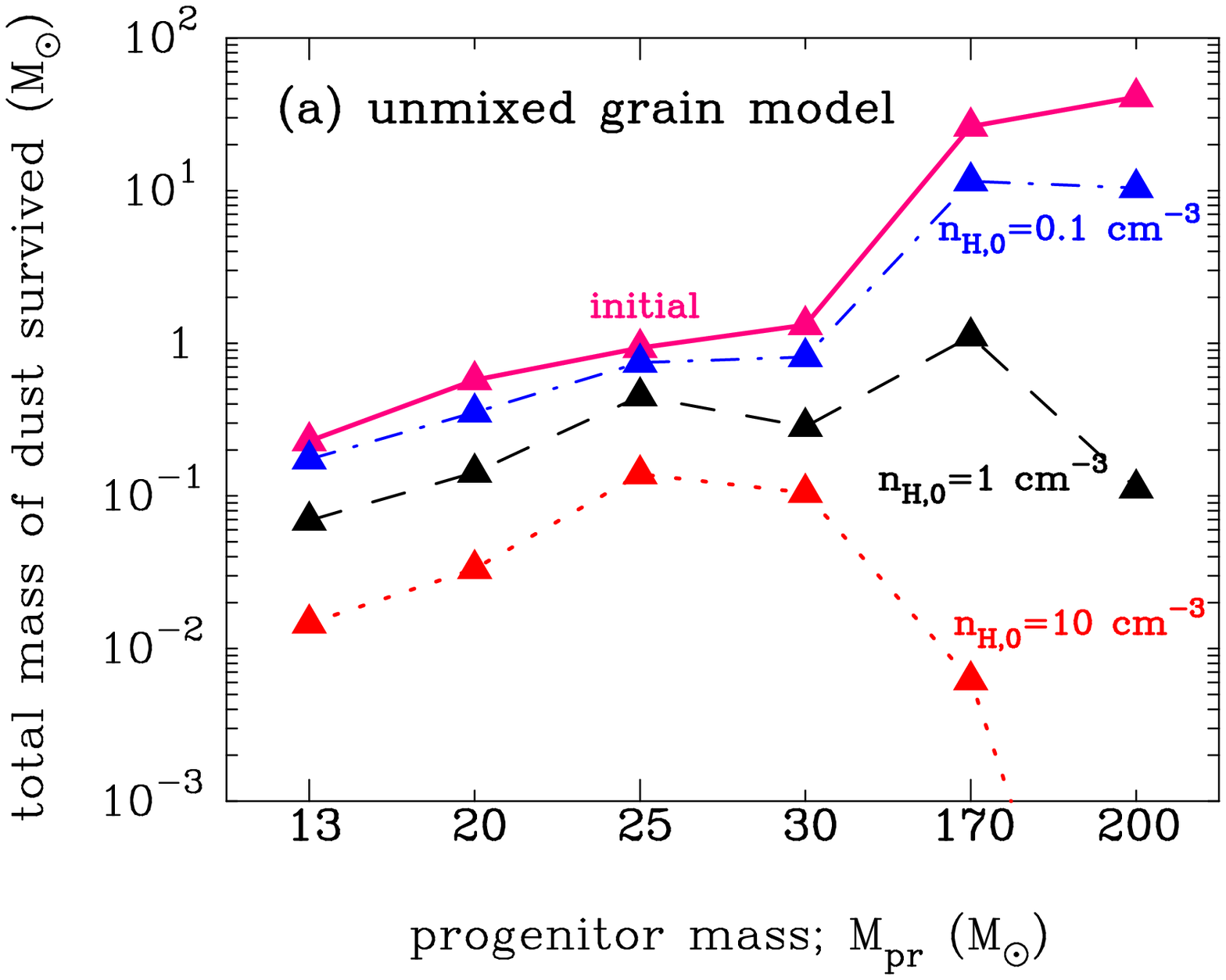}
\plotone{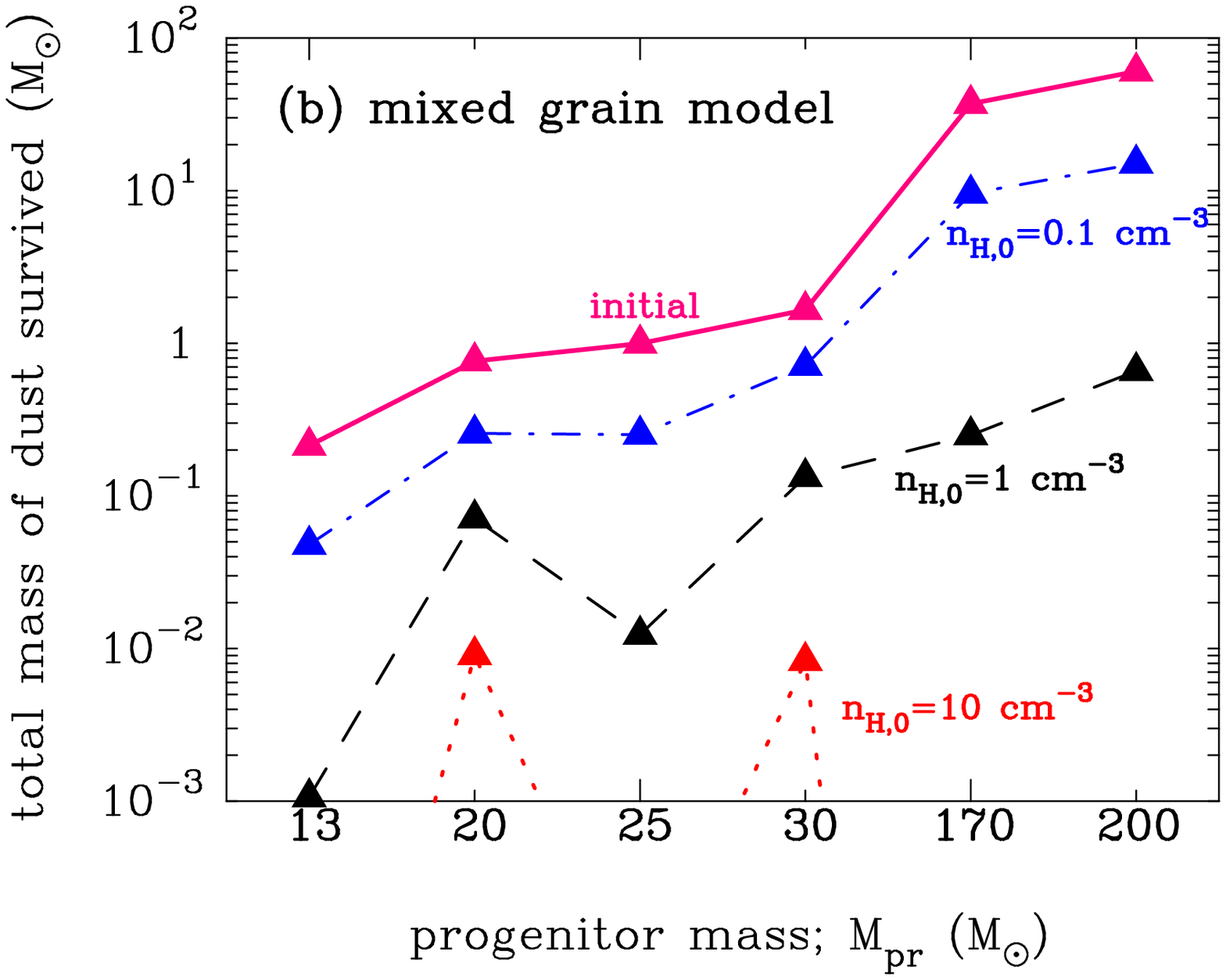}
\caption{
 The total mass of survived dust for the various progenitor mass and the 
 gas density in the ISM (a) for the unmixed grain model and (b) for the 
 mixed grain model.
 The results for $n_{\rm H,0}=$ 0.1, 1, and 10 cm$^{-3}$ are connected
 by the dot-dashed, dashed, and dotted lines, respectively.
 The solid lines are for the initial total mass of dust at the time of
 dust formation.
\textit{See the electronic edition of the Journal for a color version 
of this figure.}\label{fig10}}
\end{figure}

\clearpage

\begin{figure}
\epsscale{0.7}
\plotone{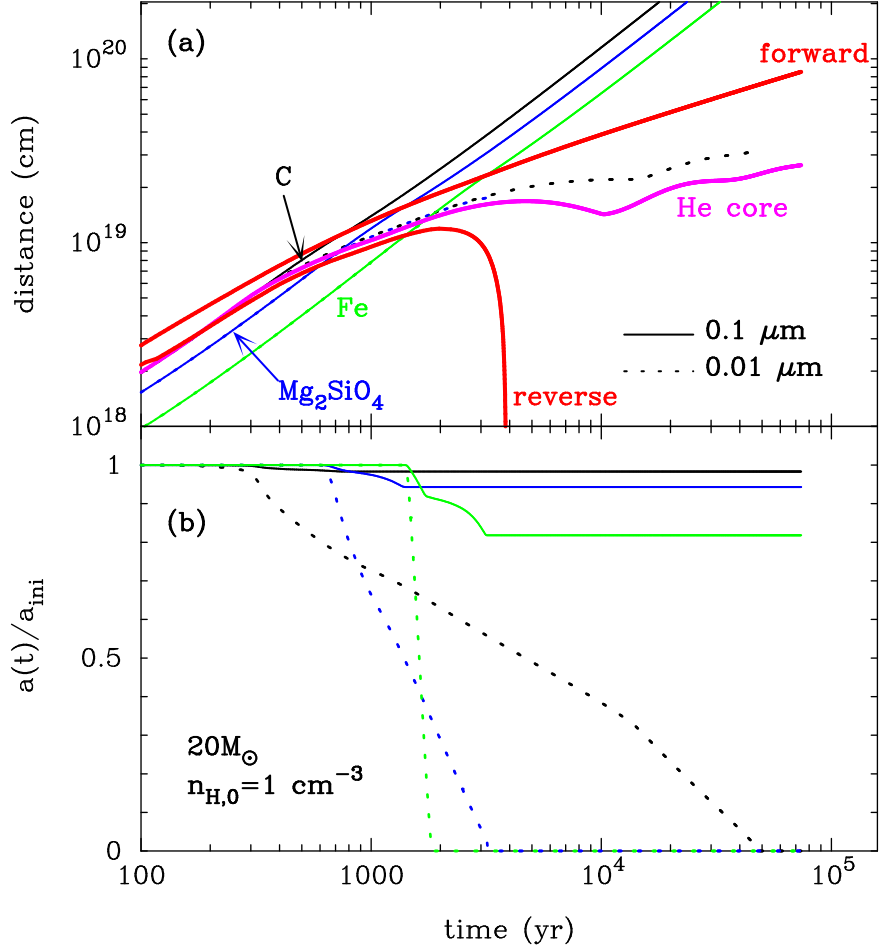}
\caption{
 The time evolutions of (a) the positions and (b) the ratios of size to 
 the initial size of dust grains within the SNR from the SN with the
 hydrogen envelope of 0.7 $M_\odot$ and the explosion energy of
 10$^{51}$ erg for $n_{\rm H,0} = 1$ cm$^{-3}$.
 The positions of the forward and reverse shocks are depicted by the 
 thick solid curves in (a), along with the position of the surface of 
 the He core.
 Among nine dust species for the unmixed grain model, are shown C, 
 Mg$_2$SiO$_4$, and Fe grains with the initial sizes of 0.01 $\micron$ 
 (\textit{dotted lines}) and 0.1 $\micron$ (\textit{solid lines}), 
 respectively.
\textit{See the electronic edition of the Journal for a color version 
of this figure.}\label{fig7}}
\end{figure}

\clearpage

\begin{deluxetable}{lccccc}
\tablewidth{0pt}
\tablecaption{THE MAIN RESULTS OF CALCULATIONS FOR THE UNMIXED GRAIN MODEL}
\tablehead{
\colhead{$M_{\rm pr}$} & 
\colhead{} & \colhead{} & \colhead{} & 
\colhead{$t_{\rm tr}$}  & \colhead{$M_{\rm shell}^{\rm H}$} \\
\colhead{($M_{\odot}$)} &
\colhead{$\epsilon_{\rm dest}$}  & \colhead{$\epsilon_{\rm eject}$} & 
\colhead{$\epsilon_{\rm shell}$} & 
\colhead{($10^6$ yr)} & \colhead{($10^4$ $M_{\odot}$)}
}
\startdata
\multicolumn{6}{c}{$n_{\rm H,0}=0.1$ cm$^{-3}$} \\
\tableline
13 & 0.240 & 0.756 & 0.004 & 2.13 & 2.96 \\                
20 & 0.386 & 0.610 & 0.004 & 2.05 & 2.58 \\
25 & 0.198 & 0.800 & 0.002 & 2.12 & 2.89 \\
30 & 0.383 & 0.614 & 0.003 & 2.11 & 2.83 \\
170 & 0.562 & 0.438 & $<$0.001 & 4.58 & 34.4 \\
200 & 0.746 & 0.247 & 0.007 & 5.13 & 50.2 \\
\tableline
\tableline
\multicolumn{6}{c}{$n_{\rm H,0}=1$ cm$^{-3}$} \\
\tableline
13 & 0.698 & 0.256 & 0.046 & 0.814 & 2.07 \\
20 & 0.752 & 0.231 & 0.017 & 0.767 & 1.72 \\
25 & 0.517 & 0.474 & 0.009 & 0.800 & 1.97 \\
30 & 0.785 & 0.206 & 0.009 & 0.790 & 1.88 \\
170 & 0.958 & 0.040 & 0.002 & 1.81 & 25.4 \\
200 & 0.997 & 0.002 & 0.001 & 2.02 & 35.2 \\
\tableline
\tableline
\multicolumn{6}{c}{$n_{\rm H,0}=10$ cm$^{-3}$} \\
\tableline
13 & 0.936 & 0.004 & 0.060 & 0.308 & 1.31 \\
20 & 0.942 & 0.013 & 0.045 & 0.301 & 1.21 \\
25 & 0.851 & 0.103 & 0.046 & 0.310 & 1.33 \\
30 & 0.920 & 0.042 & 0.038 & 0.309 & 1.31 \\
170 & 0.999 & $<$0.001 & $<$0.001 & 0.723 & 18.0 \\
200 & 1.00  & $<$0.001 & $<$0.001 & 0.825 & 27.1 \\
\enddata
\tablecomments{
 For a given $n_{\rm H,0}$, the mass fraction of dust destroyed 
 $\epsilon_{\rm dest}$, piled up in the dense shell 
 $\epsilon_{\rm shell}$, and ejected to the ISM $\epsilon_{\rm eject}$ 
 as a function of the progenitor mass $M_{\rm pr}$ for the unmixed grain 
 model.
 The truncation time $t_{\rm tr}$ and the hydrogen mass $M^{\rm H}_{\rm
 shell}$ in the dense SN shell are given by the units of $10^6$
 yr and $10^4$ $M_\odot$, respectively 
}
\end{deluxetable}

\clearpage

\begin{deluxetable}{lccccc}
\tablewidth{0pt}
\tablecaption{THE MAIN RESULTS OF CALCULATIONS FOR THE MIXED GRAIN MODEL}
\tablehead{
\colhead{$M_{\rm pr}$} & 
\colhead{} & \colhead{} & \colhead{} & 
\colhead{$t_{\rm tr}$}  & \colhead{$M_{\rm shell}^{\rm H}$} \\
\colhead{($M_{\odot}$)} &
\colhead{$\epsilon_{\rm dest}$}  & \colhead{$\epsilon_{\rm eject}$} & 
\colhead{$\epsilon_{\rm shell}$} & 
\colhead{($10^6$ yr)} & \colhead{($10^4$ $ M_{\odot}$)}
}
\startdata
\multicolumn{6}{c}{$n_{\rm H,0}=0.1$ cm$^{-3}$} \\
\tableline
13 & 0.775 & 0.212 & 0.013 & 2.14 & 2.96 \\                
20 & 0.664 & 0.311 & 0.025 & 2.05 & 2.57 \\
25 & 0.747 & 0.245 & 0.008 & 2.10 & 2.82 \\
30 & 0.569 & 0.415 & 0.016 & 2.10 & 2.80 \\
170 & 0.742 & 0.258 & $<$0.001 & 4.55 & 33.7 \\
200 & 0.751 & 0.248 & 0.001 & 5.12 & 49.8 \\
\tableline
\tableline
\multicolumn{6}{c}{$n_{\rm H,0}=1$ cm$^{-3}$} \\
\tableline
13 & 0.995 & 0.001 & 0.004 & 0.814 & 2.07 \\
20 & 0.907 & 0.075 & 0.018 & 0.767 & 1.71 \\
25 & 0.988 & 0.009 & 0.003 & 0.800 & 1.95 \\
30 & 0.920 & 0.060 & 0.020 & 0.788 & 1.86 \\
170 & 0.993 & 0.007 & $<$0.001 & 1.81 & 24.8 \\
200 & 0.989 & 0.011 & $<$0.001 & 2.00 & 34.2 \\
\tableline
\tableline
\multicolumn{6}{c}{$n_{\rm H,0}=10$ cm$^{-3}$} \\
\tableline
13 & 1.00  & 0.    & $<$0.001 & 0.309 & 1.33 \\
20 & 0.988 & 0.002 & 0.010     & 0.301 & 1.21 \\
25 & 1.00  & 0.    & $<$0.001 & 0.316 & 1.38 \\
30 & 0.995 & $<$0.001 & 0.005 & 0.308 & 1.29 \\
170 & 1.00  & 0. & $<$0.001 & 0.726 & 17.8 \\
200 & 1.00  & 0. & 0.        & 0.817 & 26.0 \\
\enddata
\tablecomments{
 Same as Table 1, but for the mixed grain model.
}
\end{deluxetable}

\clearpage

\begin{deluxetable}{lccccccccc}
\tablewidth{0pt}
\tablecaption{THE DESTRUCTION EFFICIENCY OF EACH DUST SPECIES FOR THE
UNMIXED GRAIN MODEL}
\tablehead{
\colhead{$M_{\rm pr}$} & 
\colhead{} & \colhead{} & \colhead{} & \colhead{} & \colhead{} & 
\colhead{} & \colhead{} & \colhead{} & \colhead{} \\
\colhead{($M_{\odot}$)} &
\colhead{Fe}   & \colhead{Si} & \colhead{FeS} & \colhead{MgSiO$_3$} &
\colhead{SiO$_2$}   & \colhead{Al$_2$O$_3$} & 
\colhead{Mg$_2$SiO$_4$}   & \colhead{MgO}  & \colhead{C}
}
\startdata
\multicolumn{10}{c}{$n_{\rm H,0}=0.1$ cm$^{-3}$} \\
\tableline
13 & 0.209 & 0.257 & 0.939 & 0.892 & 0.297 
   & 1.00  & 0.349 & 0.615 & 0.048 \\                
20 & 0.210 & 0.127 & 0.881 & 0.945 & 0.518 
   & 1.00  & 0.437 & 0.440 & 0.146 \\
25 & 0.089 & 0.061 & 0.693 & 0.570 & 0.086 
   & 0.999 & 0.201 & 0.520 & 0.080 \\
30 & 0.099 & 0.077 & 0.636 & 0.901 & 0.277 
   & 0.992 & 0.543 & 0.412 & 0.299 \\
170 & 0.644 & 0.273 & 1.00  & 0.987 & 0.441
    & 1.00  & 0.776 & 0.772 & 0.284 \\
200 & 0.712 & 0.787 & 1.00  & 0.990 & 0.438
    & 1.00  & 0.865 & 0.820 & 0.210 \\
\tableline
\tableline
\multicolumn{10}{c}{$n_{\rm H,0}=1$ cm$^{-3}$} \\
\tableline
13 & 0.715 & 0.660 & 1.00  & 0.998 & 0.882 
   & 1.00  & 0.964 & 0.997 & 0.520 \\
20 & 0.661 & 0.493 & 0.999 & 0.999 & 0.926 
   & 1.00  & 0.961 & 0.925 & 0.447 \\
25 & 0.379 & 0.289 & 0.984 & 0.968 & 0.369 
   & 1.00  & 0.719 & 0.934 & 0.284 \\
30 & 0.405 & 0.349 & 0.970 & 0.997 & 0.745
   & 1.00  & 0.989 & 0.946 & 0.842 \\
170 & 0.999 & 0.872 & 1.00  & 1.00  & 0.998  
    & 1.00  & 1.00  & 1.00  & 0.985 \\
200 & 0.989 & 0.999 & 1.00  & 1.00  & 0.992
    & 1.00  & 1.00  & 1.00  & 0.996 \\
\tableline
\tableline
\multicolumn{10}{c}{$n_{\rm H,0}=10$ cm$^{-3}$} \\
\tableline
13 & 0.958 & 0.894 & 1.00  & 1.00  & 0.995
   & 1.00  & 1.00  & 1.00  & 0.893 \\
20 & 0.961 & 0.851 & 1.00  & 1.00  & 1.00 
   & 1.00  & 1.00  & 1.00  & 0.896 \\
25 & 0.861 & 0.699 & 1.00  & 1.00  & 1.00 
   & 1.00  & 0.998 & 1.00  & 0.727 \\
30 & 0.824 & 0.706 & 1.00  & 1.00  & 0.958 
   & 1.00  & 1.00  & 1.00  & 0.990 \\
170 & 1.00  & 0.999 & 1.00  & 1.00  & 1.00  
    & 1.00  & 1.00  & 1.00  & 1.00 \\
P200 & 1.00  & 1.00  & 1.00  & 1.00  & 1.00  
     & 1.00  & 1.00  & 1.00  & 1.00 \\
\enddata
\tablecomments{
 For a given $n_{\rm H,0}$, the mass fraction of dust destroyed 
 as a function of the progenitor mass $M_{\rm pr}$ for each grain
 species in the unmixed grain model.
}
\end{deluxetable}

\clearpage

\begin{deluxetable}{lccccc}
\tablewidth{0pt}
\tablecaption{THE DESTRUCTION EFFICIENCY OF EACH DUST SPECIES FOR THE 
MIXED GRAIN MODEL}
\tablehead{
\colhead{$M_{\rm pr}$} & 
\colhead{} & \colhead{} & 
\colhead{} & \colhead{} & \colhead{} \\
\colhead{($M_{\odot}$)} & 
\colhead{Al$_2$O$_3$}   & \colhead{MgSiO$_3$} &
\colhead{Mg$_2$SiO$_4$} & \colhead{SiO$_2$}  & \colhead{Fe$_3$O$_4$}
}
\startdata
\multicolumn{6}{c}{$n_{\rm H,0}=0.1$ cm$^{-3}$} \\
\tableline
13 & 1.00  & 0.862 & 0.868 & 0.676 & 0.788 \\                
20 & 1.00  & 0.913 & 0.788 & 0.518 & 0.992 \\
25 & 1.00  & 0.910 & 0.877 & 0.596 & 0.999 \\
30 & 0.997 & 0.802 & 0.620 & 0.461 & 0.999 \\
170 & 1.00  & 0.961 & 0.957 & 0.585 & 1.00 \\
200 & 1.00  & 0.984 & 0.995 & 0.615 & 0.999 \\
\tableline
\tableline
\multicolumn{6}{c}{$n_{\rm H,0}=1$ cm$^{-3}$} \\
\tableline
13 & 1.00 & 0.999 & 0.999 & 0.988 & 0.999 \\
20 & 1.00 & 0.996 & 0.977 & 0.850 & 1.00  \\
25 & 1.00 & 1.00  & 1.00  & 0.976 & 1.00  \\
30 & 1.00 & 0.994 & 0.951 & 0.875 & 1.00  \\
170 & 1.00  & 1.00  & 1.00  & 0.989 & 1.00 \\
200 & 1.00  & 1.00  & 1.00  & 0.983 & 1.00  \\
\tableline
\tableline
\multicolumn{6}{c}{$n_{\rm H,0}=10$ cm$^{-3}$} \\
\tableline
13 & 1.00 & 1.00 & 1.00 & 0.999 & 1.00 \\
20 & 1.00 & 1.00 & 1.00 & 0.980 & 1.00 \\
25 & 1.00 & 1.00 & 1.00 & 1.00  & 1.00 \\
30 & 1.00 & 1.00 & 0.996 & 0.976 & 1.00 \\
170 & 1.00  & 1.00  & 1.00  & 1.00  & 1.00 \\
P200 & 1.00  & 1.00  & 1.00  & 1.00  & 1.00 \\
\enddata
\tablecomments{
 Same as Table 3, but for each grain species in the mixed grain model.
}
\end{deluxetable}

\clearpage

\begin{deluxetable}{lccccc}
\tablewidth{0pt}
\tablecaption{ELEMENTAL ABUNDANCES IN SN SHELL}
\tablehead{
\colhead{$M_{\rm pr}$} & 
\colhead{} & \colhead{} & 
\colhead{} & \colhead{} & \colhead{} \\
\colhead{($M_{\odot}$)} & 
\colhead{[Fe/H]}   & \colhead{[C/Fe]}  &  
\colhead{[O/Fe]}   & \colhead{[Mg/Fe]} & 
\colhead{[Si/Fe]}  
}
\startdata
\multicolumn{6}{c}{$n_{\rm H,0}=0.1$ cm$^{-3}$} \\
\tableline
13 & $-6.43$ & $-0.274$ & $-0.699$ & $-0.230$ & $1.92$ \\
20 & $-5.20$ & $0.117$  & $-0.595$ & $0.034$  & $0.410$ \\
25 & $-5.90$ & $1.11$   & $-1.42$  & $-0.500$ & $-0.552$ \\
30 & $-5.56$ & $0.566$  & $-0.043$ & $0.739$  & $0.866$ \\
\tableline
\tableline
\multicolumn{6}{c}{$n_{\rm H,0}=1$ cm$^{-3}$} \\
\tableline
13 & $-5.15$ & $1.11$  & $-0.555$ & $-0.459$ & $1.01$ \\
20 & $-5.53$ & $0.992$ & $0.585$  & $1.16$   & $1.87$ \\
25 & $-5.23$ & $1.09$  & $-0.412$ & $0.407$  & $0.989$ \\
30 & $-5.11$ & $0.797$ & $0.242$  & $1.09$   & $1.26$  \\
\tableline
\tableline
\multicolumn{6}{c}{$n_{\rm H,0}=10$ cm$^{-3}$} \\
\tableline
13 & $-4.13$ & $0.284$  & $-2.54$ & $-3.89$ & $0.599$ \\
20 & $-4.92$ & $0.946$  & $-2.15$ & $-1.80$ & $2.14$ \\
25 & $-5.10$ & $1.60$   & $0.122$ & $0.232$ & $2.34$ \\
30 & $-5.11$ & $-0.207$ & $0.375$ & $-1.23$ & $2.66$ \\
\enddata
\tablecomments{
 [Fe/H] and abundances of C, O, Mg, and Si relative to Fe in the shell
 of primordial Type II SNRs for given $M_{\rm pr}$ and $n_{\rm H,0}$
 with the unmixed grain model.
}
\end{deluxetable}

%

\end{document}